\documentclass[aps,prd,reprint,superscriptaddress,nofootinbib]{revtex4-2}
\usepackage{amsmath,amsfonts,amssymb,bm}
\usepackage{graphicx}
\usepackage{bbold}
\usepackage[normalem]{ulem}
\usepackage[dvipsnames,usenames]{xcolor}
\usepackage[colorlinks,citecolor=blue]{hyperref}
\usepackage{orcidlink}

\providecommand{\apjs}{Astrophys. J. Suppl.}
\providecommand{\jcap}{J. Cosm. Astropart. Phys.}

\begin{document}

\title{Occurrence of fast neutrino flavor conversions in QCD phase-transition supernovae}

\author{Zewei~Xiong\orcidlink{0000-0002-2385-6771}}
\email[Corresponding Author: ]{z.xiong@gsi.de}
\affiliation{GSI Helmholtzzentrum {f\"ur} Schwerionenforschung, Planckstra{\ss}e 1, D-64291 Darmstadt, Germany}

\author{Meng-Ru~Wu\orcidlink{0000-0003-4960-8706}}
\email{mwu@as.edu.tw}
\affiliation{Institute of Physics, Academia Sinica, Taipei 115201, Taiwan}
\affiliation{Institute of Astronomy and Astrophysics, Academia Sinica, Taipei 106319, Taiwan}
\affiliation{Physics Division, National Center for Theoretical Sciences, Taipei 106319, Taiwan}

\author{Noshad~Khosravi~Largani\orcidlink{0000-0003-1551-0508}}
\affiliation{Institute of Theoretical Physics, University of Wroclaw, Pl. M. Borna 9, 50-204 Wroclaw, Poland}

\author{Tobias~Fischer\orcidlink{0000-0003-2479-344X}}
\affiliation{Institute of Theoretical Physics, Wrocław University of Science and Technology, Wybrzeże Wyspiańskiego 27, 50-370 Wrocław, Poland}
\affiliation{Research Center for Computational Physics and Data Processing, Institute of Physics, Silesian University in Opava, Bezručovo nám. 13, CZ-746-01 Opava, Czech Republic}

\author{Gabriel~Mart{\'i}nez-Pinedo\orcidlink{0000-0002-3825-0131}}
\affiliation{GSI Helmholtzzentrum {f\"ur} Schwerionenforschung, Planckstra{\ss}e 1, D-64291 Darmstadt, Germany}
\affiliation{Institut f{\"u}r Kernphysik (Theoriezentrum), Fachbereich Physik, Technische Universit{\"a}t Darmstadt, Schlossgartenstra{\ss}e 2, D-64289 Darmstadt, Germany}

\date{\today}

\begin{abstract}
Core-collapse supernovae undergoing a first-order quantum chromodynamics (QCD) phase transition experience the collapse of the central proto-neutron star that leads to a second bounce.
This event is accompanied by the release of a second neutrino burst.
Unlike the first stellar core bounce neutrino burst, which consists exclusively of electron neutrinos, the second burst is dominated by electron antineutrinos.
Such a condition makes QCD supernovae an ideal site for the occurrence of fast neutrino flavor conversion (FFC), which can lead to rapid flavor equilibration and significantly impact the related neutrino signal.
In this work, we perform a detailed analysis of the conditions for fast flavor instability (FFI) around and after the second neutrino burst in QCD phase transition supernova models launched from 25~$M_\odot$ and 40~$M_\odot$ progenitor models.
We evaluate the relevant instability criteria and find two major phases of FFC.
The first phase is closely associated with the collapse and the rapidly expanding shock wave, which is a direct consequence of the proto-neutron star collapse due to the phase transition. 
The second phase takes place a few milliseconds later when electron degeneracy is restored near the proto-neutron star surface.
We also characterize the growth rate of FFI and estimate its impact on the evolution of the neutrino flavor content.
The potential observational consequences on neutrino signals are evaluated by comparing a scenario assuming complete flavor equipartition with other scenarios without FFC.
Finally, we investigate how FFC may influences $r$-process nucleosynthesis associated with QCD phase transition driven supernova explosions.
\end{abstract}

\maketitle
\graphicspath{{./}{figures/}}

\section{Introduction}
\label{sec:introduction}
Stars with zero age main sequence masses exceeding approximately 8--9~M$_\odot$ are the progenitors of core-collapse supernovae (SN), which undergo stellar core collapse at the end of their lives.
The collapse halts with the core bounce that leads to the formation of a hydrodynamic shock wave, when normal nuclear matter density is reached.
This bounce shock expands rapidly but suffers from energy losses owing to the dissociation of nuclei in infalling material, from the still collapsing outer layers of the stellar core, and neutrino emission.
The latter are related to the shock propagation across the neutrinospheres of last scattering, which releases a burst of electron neutrinos on the order of 5--25~ms after core bounce.
These energy losses turn the expanding bounce shock into an accretion front.
The further evolution of this shock is driven by neutrino heating and cooling in the post shock layers.
The eventual revival of the stalled bounce shock, due to the liberation of energy from the interior of the proto-neutron star (PNS), which formed at core bounce, on the order of several $10^{53}$~erg of gravitational binding energy gain, is the key subject of the supernova problem, i.e. the central engine that results in massive star explosions.
Pertaining to the question of how to revive the stalled shock, multiple mechanisms have been proposed, including the standard delayed-neutrino heating mechanism~\cite{colgate1966hydrodynamic,bethe1985revival} and the magneto-rotational SN~\cite{leblanc1970numerical}.
Besides these, an additional mechanism has been proposed in Refs.~\cite{takahara1988supernova,gentile1993qcd,sagert2009signals}, which involves a first-order quantum chromodynamics (QCD) phase transition from hadronic matter to deconfined quark matter.

The transition from hadronic to quark matter at supra-nuclear density leads to the softening of the equations of state (EOS).
As a result, the PNS core undergoes a supersonic collapse and a second shock is launched when the quark core bounces back.
A large amount of latent heat is released during the phase transition and can drive a successful SN explosion once the second shock takes over the standing accretion shock from stellar core bounce, leaving behind a significantly more compact PNS with a quark core.
Numerous simulations for QCD phase-transition SN have been performed in recent years \cite{fischer2018quark,fischer2020corecollapse,zha2020gravitationalwave,fischer2021qcd,jakobus2022role,kuroda2022corecollapse,zha2022impact,jakobus2023gravitational,largani2024constraining}, which exhibit various features different from the standard neutrino-heating SN.

One important consequence is that the conditions in QCD phase transition driven SN explosions, i.e. rapidly expanding matter featuring a large neutron excess followed by the slower high-entropy neutrino-driven wind (NDW), enable the rapid neutron capture process ($r$ process) to produce elements beyond the second $r$-process peak (atomic number $Z\sim 50$).
Traditionally the $r$ process was associated with the NDW in canonical PNS that result from neutrino-driven SN explosions \cite{woosley1994rprocess,witti1994nucleosynthesis1,takahashi1994nucleosynthesis,qian1996nucleosynthesis}. 
Improved microphysics, including updated weak rates employed in 
self-consistent SN simulations, which are based on three-flavor Boltzmann neutrino transport, yield the nucleosynthesis of light neutron-capture elements with a sharp drop of the abundances beyond atomic number $Z>45$
\cite{stockinger2020threedimensional,bollig2021selfconsistent,witt2021postexplosion,wang2023neutrinodriven,fischer2024neutrinos}.  
In contrast, the QCD SN gives a wider variety of nucleosynthesis conditions of the material ejected at different phases \cite{nishimura2012nucleosynthesis,fischer2020corecollapse,jakobus2022role}.
Three types of ejecta have been classified: direct ejecta, intermediate ejecta, and NDW.
The direct ejecta follow the second shock and expand at relativistic velocities.
The expansion is fast enough such that weak processes can only partly change the composition, i.e. reducing the neutron excess. 
The material that fails to be directly ejected, falls back toward the PNS at an intermediate time stage.
Before accretion onto the PNS surface, the neutrino energy deposition renders their re-ejection. 
Simultaneously, the very same process also reduces the neutron richness of the intermediate ejecta, i.e. electron neutrino and antineutrino absorptions on neutrons and protons.
In the last phase, NDW is launched from the PNS surface.  
With a more compact PNS, the typical entropy of NDW at a timescale of seconds is much higher than that of the canonical wind, since the NDW entropy depends sensitively on the mass-to-radius ratio of the nascent PNS~\cite{qian1996nucleosynthesis}. 
The large neutron excess in direct ejecta and the high entropy NDW create suitable conditions for the strong $r$-process to take place, with the production of nuclei beyond the third $r$-process peak at mass number $A\sim 190$, including the actinides.

Another prominent feature of QCD phase transition driven SN is a millisecond burst neutrino signature associated with the propagation of the second shock across the neutrinospheres of last scattering \cite{fischer2018quark,largani2024constraining,jakobus2022role,kuroda2022corecollapse,zha2022impact}.
While this millisecond burst is mainly dominated by $\bar\nu_e$, other species are also largely present, making it an ideal detection target for current and future generations of large-scale neutrino detectors capable of detecting different flavors~\cite{dasgupta2010detecting}.
The expected neutrino event rate during the millisecond burst could be one order of magnitude larger than that before or after the burst.
A detection of such a burst would not only reveal the occurrence of phase transition in SN cores, but also provide an opportunity to probe other aspects through multimessenger observations, including, e.g., the improved determination of the position of galactic SN with triangulation using the pinpointed timing of the burst~\cite{pitik2022exploiting}, and the correlated analysis between gravitational waves and the modulation of neutrino events~\cite{zha2020gravitationalwave,zha2022impact,lin2024detectability}.
There can be interesting implications of QCD phase transition in other astrophysical environments as well (e.g., \cite{bauswein2019identifying,most2019signatures,chan2025distinct}).

To fully understand the aforementioned features of QCD SN, an accurate treatment of neutrino radiation transport is essential.
In particular, the antineutrino-dominated burst and the associated neutrino signal as well as the nuclear composition in the ejecta above the PNS all depend on the modeling of neutrino transport. 
However, despite the implementation of six-species Boltzmann neutrino transport, there is one incompletely understood but important aspect of neutrino physics in current SN simulations, namely the so-called fast flavor conversion (FFC), which are triggered by fast flavor instability (FFI); for recent reviews, see Refs.~\cite{volpe2024neutrinos,johns2025neutrino}.
When FFC occur, neutrinos propagating in different directions and carrying opposite electron-minus-heavy-flavor lepton number ($\nu$ELN) can transform their flavors at an extremely rapid rate within a timescale of less than a nanosecond. 
While FFI was only identified in very limited domains in spherically symmetric neutrino-driven SN simulations~\cite{morinaga2020fast}, its more general existence in various regions was found in multidimensional SN and neutron star merger simulations~\cite{abbar2019occurrence,delfanazari2019linear,delfanazari2020fast,nagakura2019fastpairwise,morinaga2020fast,abbar2020fast,glas2020fast,nagakura2021where,abbar2021characteristics,harada2022prospects,akaho2024collisional,wu2017imprints,george2020fast,li2021neutrino,just2022fast,fernandez2022fast,mukhopadhyay2024time,nagakura2025neutrino}.
The main reason that limits the FFI condition in spherically symmetric {\em canonical} SN~\cite{tamborra2017flavordependent,xiong2023evolution,liu2023universality} is the dominant emission of $\nu_e$ over $\bar\nu_e$ due to the deleptonization.
It is natural to wonder whether the $\bar\nu_e$-dominated emission associated with the release of the second neutrino burst in QCD phase transition driven SN can relax this constraint, leading to general presence of FFI in such events.
This motivates the investigation of the FFI characteristics and their implications in QCD SNe.

In this work, we examine the occurrence condition of FFI in a QCD SN model provided by~\cite{largani2024constraining}.
We use this model as an example to present the basic features of the FFI during the first few tens of milliseconds after the second bounce, associated with the formation of the second shock as a consequence of the first-order QCD phase transition and subsequent PNS collapse, and explore their potential impact on the neutrino signal and nucleosynthesis in the direct ejecta.
This paper is organized as follows.
We describe the evolution of key thermal quantities in the QCD phase-transition SN in Sec.~\ref{sec:simulation} and relevant aspects of the neutrino emission and transport in Sec.~\ref{sec:neutrinos}.
We present detailed properties associated with the FFI and discuss its potential implications in Sec.~\ref{sec:FFI}.
Finally, we summarize our conclusions in Sec.~\ref{sec:discussion}.
We adopt natural units with $\hbar=c=k_{\rm B}=1$ throughout the paper.

\section{Neutrino-radiation hydrodynamics simulations}
\label{sec:simulation}
We perform a self-consistent simulation of a SN explosion that features an early first-order QCD phase transition. 
In this section, we review the SN model and recap the evolution toward and across the phase transition.

\subsection{Core-collapse supernova model}
The SN simulation is performed with \textsc{agile-boltztran}, which is based on spherically symmetric general relativistic neutrino radiation hydrodynamics in comoving coordinates \cite{mezzacappa1993type,mezzacappa1993stellar,mezzacappa1993numerical,liebendorfer2001conservative,liebendorfer2004finite}, including six-species Boltzmann neutrino transport assuming ultra-relativistic particles~\cite{lindquist1966relativistic,fischer2020muonization}.
The equations of radiation hydrodynamics are solved on an adaptive baryon mass mesh~\cite{liebendorfer2002adaptive,fischer2010protoneutron}, for which 207 radial grid points are used for the current simulation.
The Boltzmann neutrino transport equation is solved using the discrete-ordinate method.
The neutrino phase space is discretized with respect to the lateral momentum angles, $\cos\theta\in\{-1,+1\}$, using six-point Gauss quadrature, and to the neutrino energy, $E\in\{3,300\}~{\rm MeV}$, using 24 bins following the setup of Refs.~\cite{mezzacappa1993numerical,bruenn1985stellar}.
The set of non-muonic weak interactions considered here can be found in Table~I of Ref.~\cite{largani2024constraining}, with the references given therein.
Because muonic weak processes are omitted here, muon and tau flavors are indistinguishable and are collectively denoted by ``$x$'', even if they are treated separately in \textsc{agile-boltztran}. 
No neutrino oscillations are taken into consideration in the SN model. 

\textsc{agile-boltztran} has a flexible EOS module that can handle a variety of nuclear matter EOS~\cite{lattimer1991generalized,Shen98,hempel2012new,hempel2012new,steiner2013corecollapse}, including the presence of hadronic strange resonances~\cite{Fischer2025JCAP01}, and those including exotic phases at high density and temperature~\cite{sagert2009signals,Fischer11,fischer2018quark,Fischer:2021}.
The sub-saturation density EOS is based on the modified nuclear statistical equilibrium model of Ref.~\cite{hempel2010statistical}, featuring several thousands of nuclear species and including an approximate treatment of excited states. 
For the low-density and low-temperature regime with $T<0.45$~MeV, a silicon-gas approximation is used for the baryons, including Coulomb contributions. 
Further contributions from electrons, positrons and photons are also added at all conditions based on Refs.~\cite{timmes1999accuracy,timmes2000accuracy}. 

The hadronic-quark hybrid EOS used in this study is RDF-1.9~\cite{bastian2021phenomenological}, featuring the DD2 hadronic EOS based on the relativistic mean field parametrization~\cite{typel2010composition}, with density-dependent nucleon-meson couplings. 
At high density and temperatures, RDF-1.9 takes into account a first-order phase transition to deconfined quark matter, based on a phase transition construction following the Maxwell approach. 
The quark matter phase is modeled using the string-flip EOS, derived from a certain truncation from the QCD gap equations and expressed in the relativistic density functional (RDF) framework~\cite{kaltenborn2017quarknuclear}.
Specifically, the RDF-1.9 model features a low onset density, compared to other RDF models from the catalog of Ref.~\cite{bastian2021phenomenological}.
For $T=0$, we have an onset density for quark matter of $\rho_{\rm onset}=4.6\times 10^{14}~\mathrm{g~cm}^{-3}$, and for entropy per particle $s=3$ we have $\rho_{\rm onset}=3.3\times 10^{14}~\mathrm{g~cm}^{-3}$~\cite{largani2024constraining}.

The simulation discussed below is launched from the stellar progenitor star with zero-age main sequence mass of $25$~M$_\odot$ and solar metallicity, from the stellar evolution series of Ref.~\cite{rauscher2002nucleosynthesis}.
Henceforth, the present investigation of the selected model is denoted as {\tt s25a28} RDF-1.9.
The simulation extends to a post-bounce time of $t_{\rm pb}=378$~ms and leads to a successful explosion driven by the first-order QCD phase transition with a diagnostic breakout energy of $1.45\times 10^{51}$~erg.
With this model, the phase transition occurs earlier than for all other models with different RDF EOS from the same progenitor star (further details can be found in Ref.~\cite{largani2024constraining}).
The second bounce following the phase transition occurs at $t_{\rm pb}=343$~ms, which we denote as the post-second-bounce time $t_{\rm p2b}=0$.
The PNS has an enclosed mass of about $M_{\rm PNS}=2.0$~M$_\odot$ with a quark matter core of $M_{\rm quark}=1.72$~M$_\odot$. 
This is below the maximum mass that can be supported by the corresponding EOS at the phase-transition onset to avoid collapsing into a black hole (see Table~2 in Ref.~\cite{largani2024constraining}).

\begin{figure*}[!hbt]
\centering
\includegraphics[width=\columnwidth]{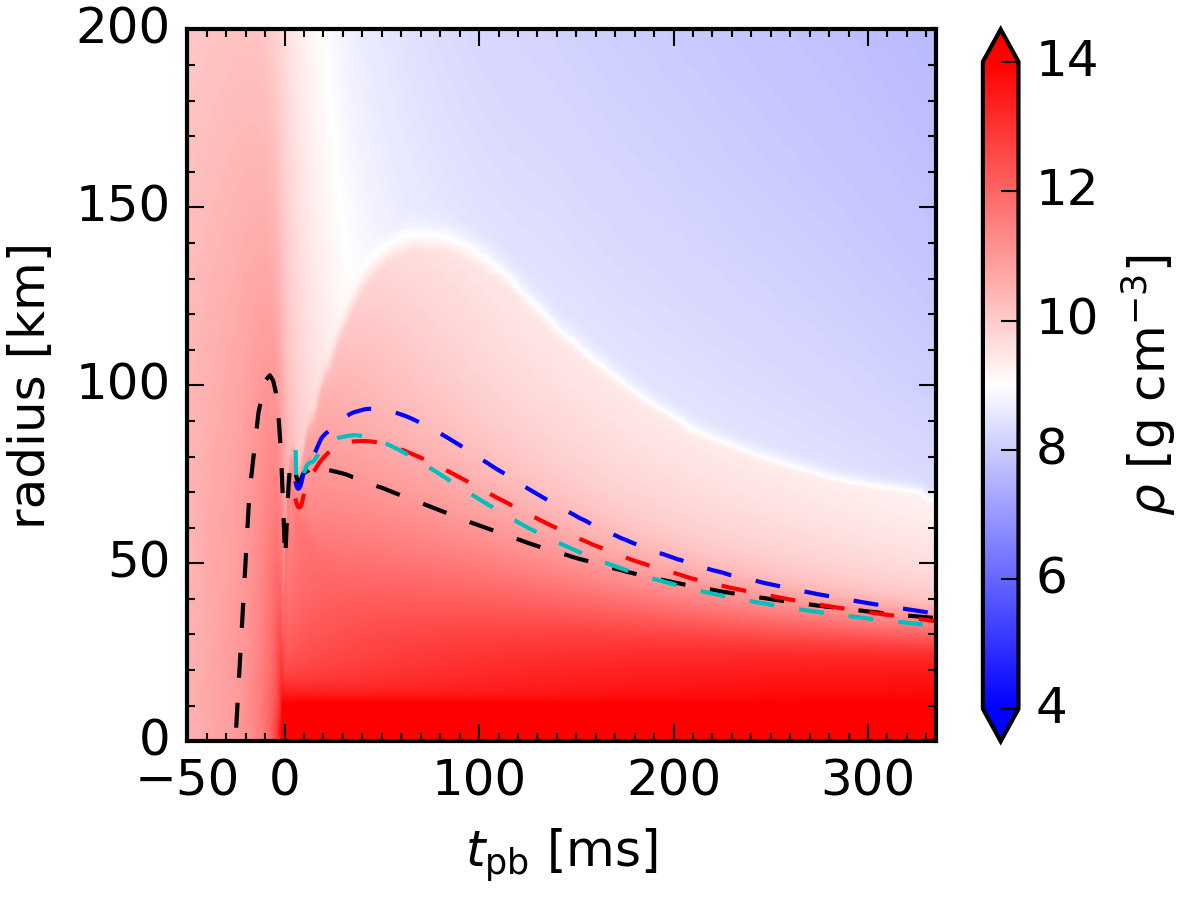}
\includegraphics[width=\columnwidth]{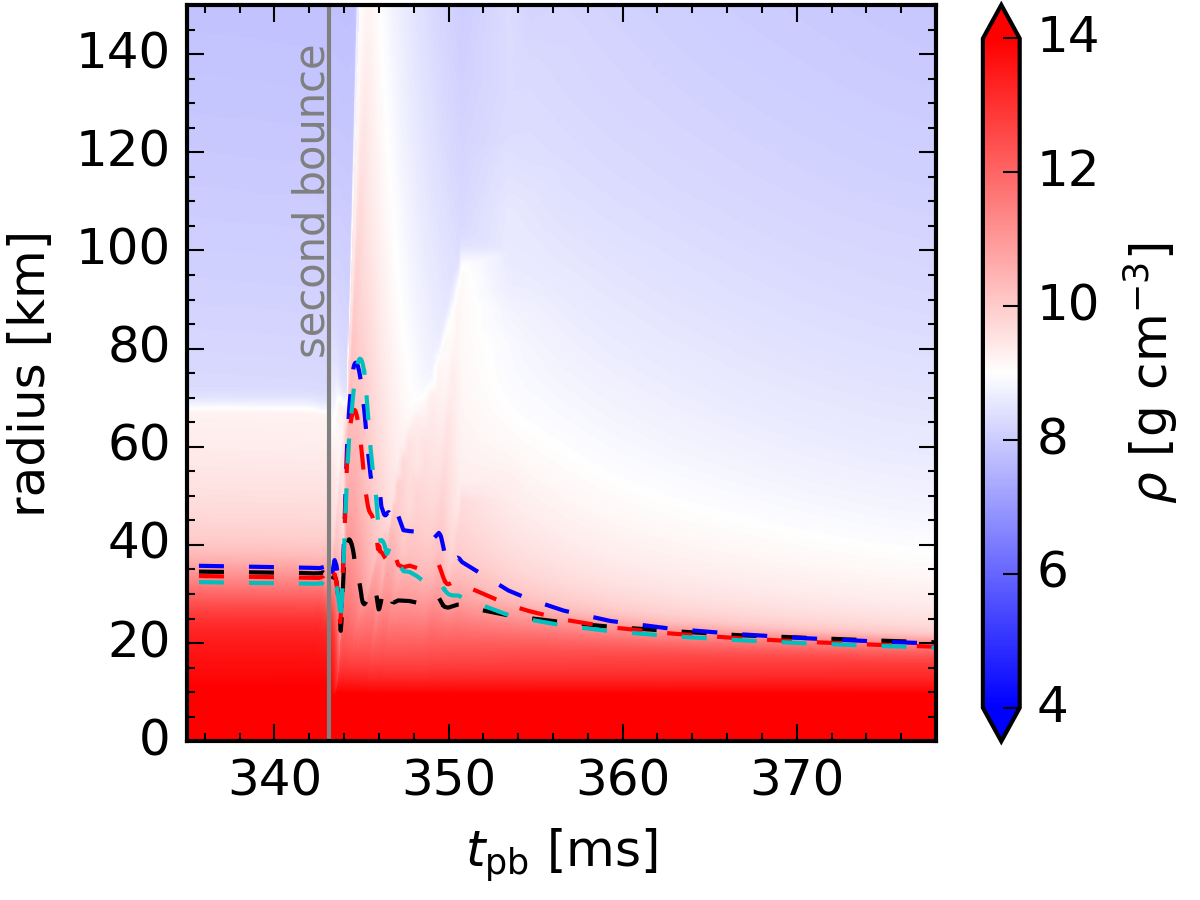}
\llap{\parbox[b]{8.9in}{\small (a)\\\rule{0ex}{2.2in}}}
\llap{\parbox[b]{2.1in}{\small (b)\\\rule{0ex}{2.2in}}}
\includegraphics[width=\columnwidth]{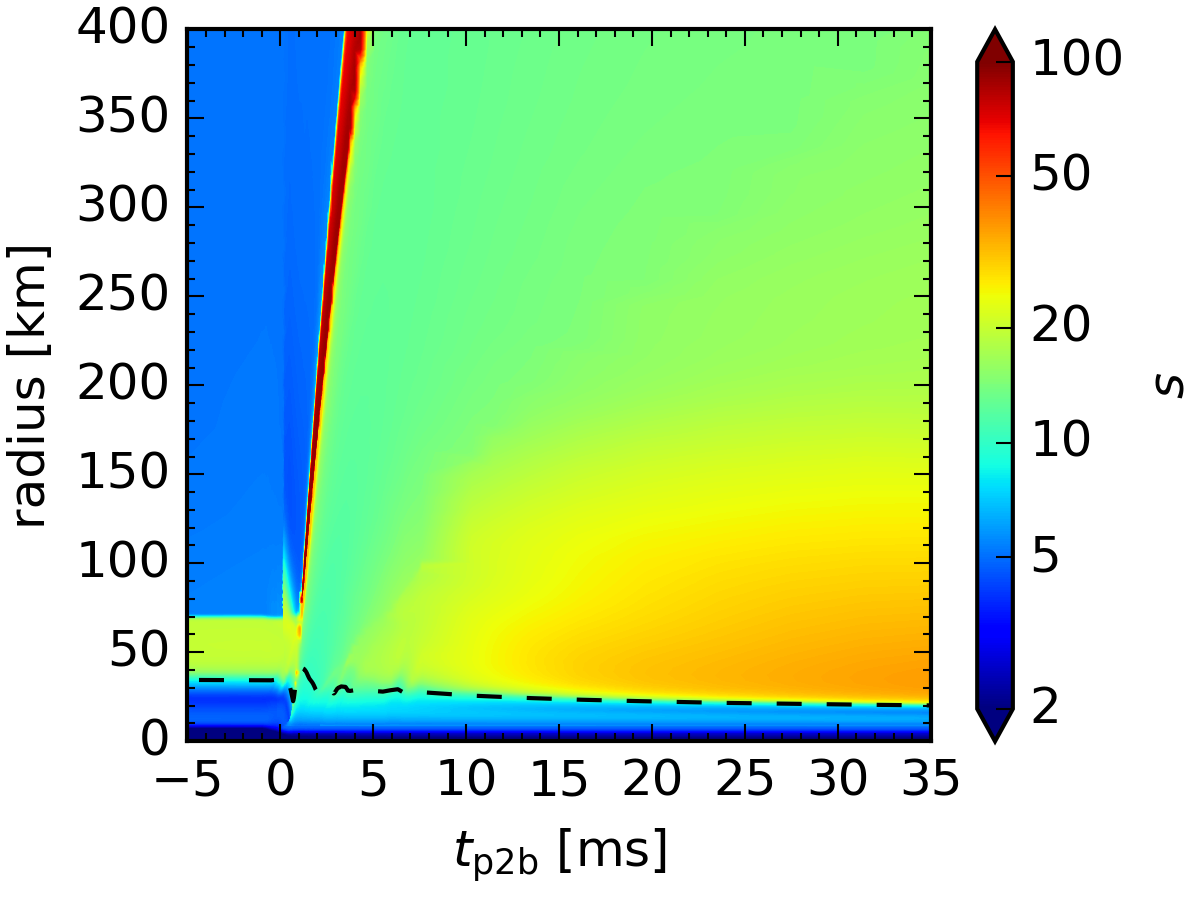}
\includegraphics[width=\columnwidth]{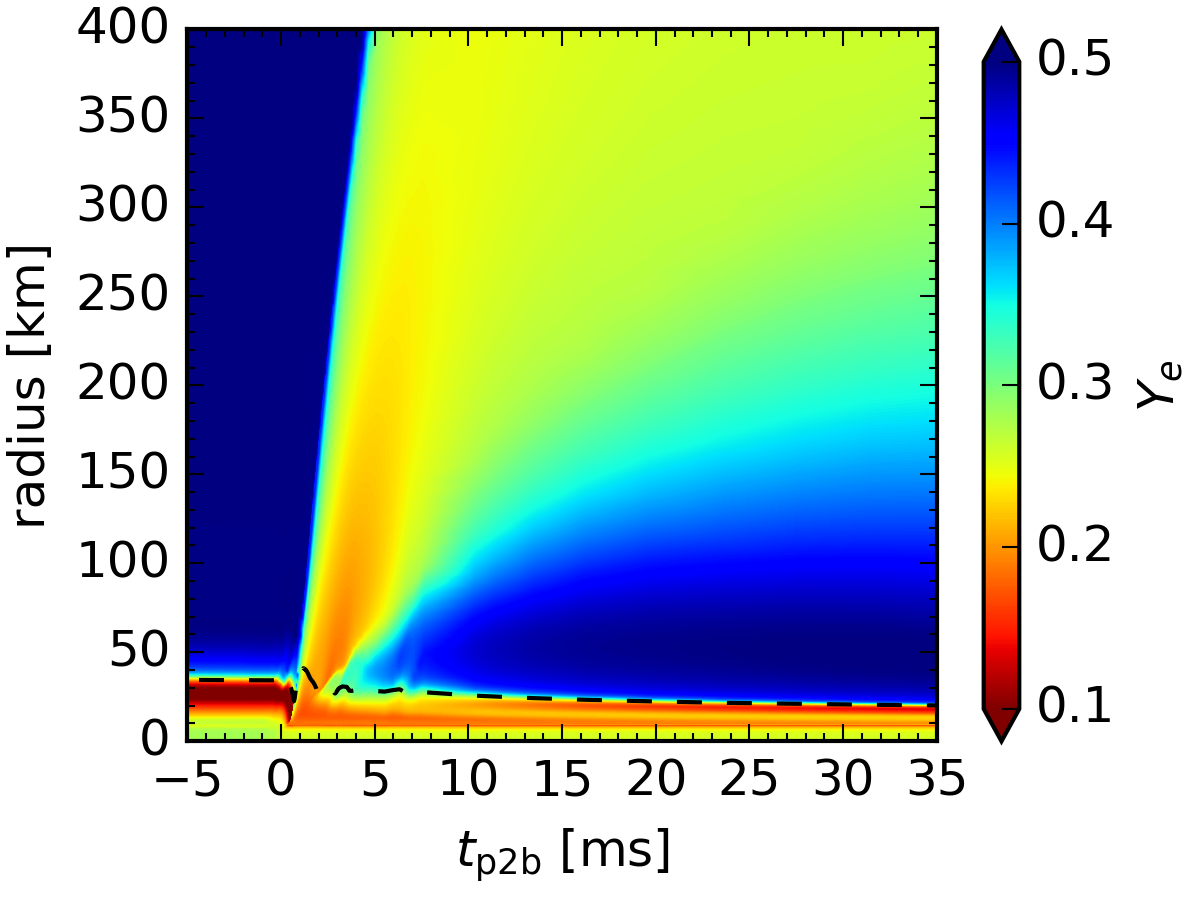}
\llap{\parbox[b]{8.9in}{\small (c)\\\rule{0ex}{2.2in}}}
\llap{\parbox[b]{2.1in}{\small (d)\\\rule{0ex}{2.2in}}}
\caption{Radial evolution of selected quantities for the reference SN model {\tt s25a28}~RDF-1.9, showing the rest-mass density $\rho$ (a--b), the entropy per particle $s$ (c), and the electron fraction $Y_e$ (d), with respect to the post bounce time $t_{\rm pb}$ (top panels) and post-second-bounce time $t_{\rm p2b}$ (bottom panels). 
Black, blue, red, and cyan dashed curves in (a) and (b) show the radii of PNS, $\nu_e$-, $\bar\nu_e$-, and $\nu_x$-spheres, respectively.}
\label{fig:2d_thermo}
\end{figure*}

\begin{figure*}[!hbt]
\centering
\includegraphics[width=0.97\textwidth]{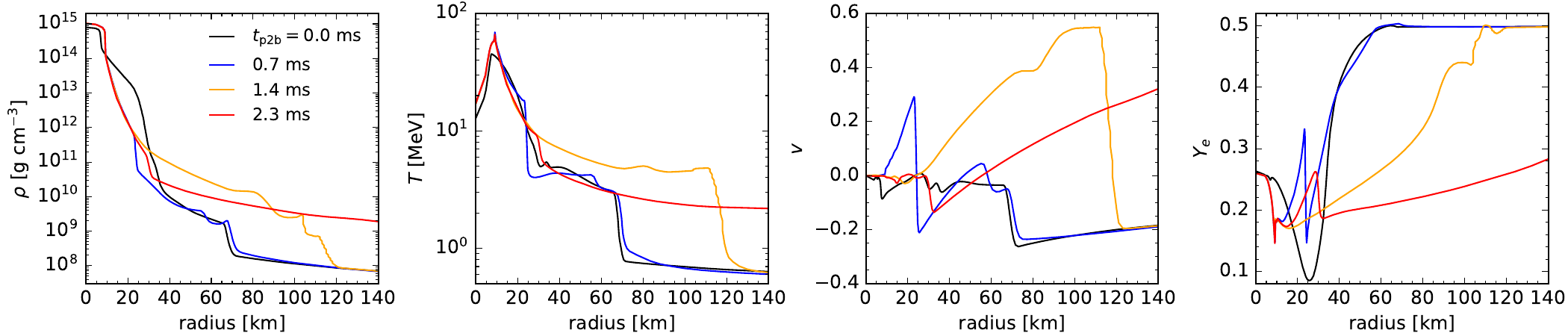}
\llap{\parbox[b]{12.5in}{\small (a)\\\rule{0ex}{1.2in}}}
\llap{\parbox[b]{9.2in}{\small (b)\\\rule{0ex}{1.2in}}}
\llap{\parbox[b]{6.0in}{\small (c)\\\rule{0ex}{1.2in}}}
\llap{\parbox[b]{2.7in}{\small (d)\\\rule{0ex}{1.2in}}}
\includegraphics[width=0.97\textwidth]{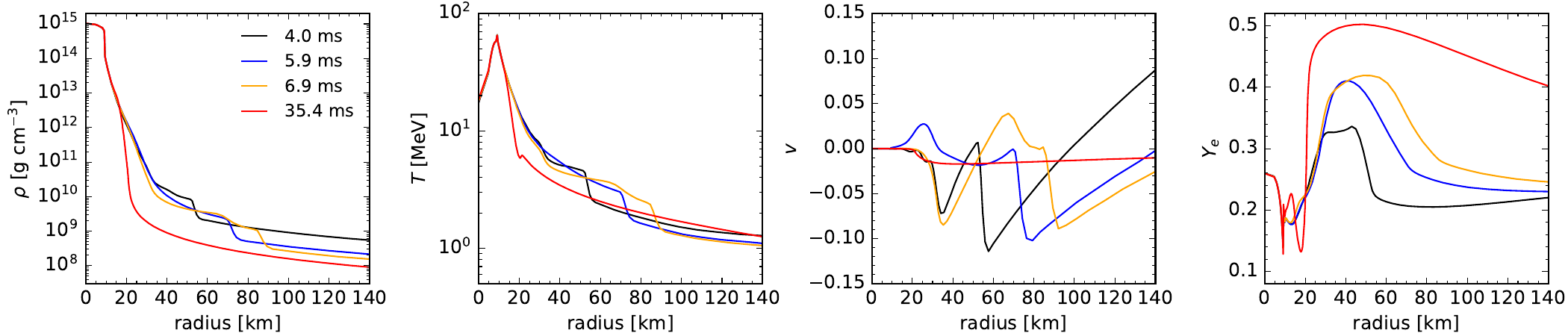}
\llap{\parbox[b]{12.5in}{\small (e)\\\rule{0ex}{1.2in}}}
\llap{\parbox[b]{9.2in}{\small (f)\\\rule{0ex}{1.2in}}}
\llap{\parbox[b]{6.0in}{\small (g)\\\rule{0ex}{1.2in}}}
\llap{\parbox[b]{2.7in}{\small (h)\\\rule{0ex}{1.2in}}}
\caption{ Radial profiles of rest-mass density, $\rho$, temperature, $T$ , radial velocity, $v$, and electron fraction, $Y_e$, at different post-second-bounce times, $t_{\rm p2b}$, showing the early (a--d) and the later evolution (e--h).}
\label{fig:rt}
\end{figure*}

\subsection{Evolution across QCD phase transition}
The evolution of the rest-mass density $\rho$ is shown in Figs.~\ref{fig:2d_thermo}(a)~and~(b), including first and second bounces.
The black dashed curve marks the PNS surface at $\rho=10^{11}~\mathrm{g~cm}^{-3}$. 
Other dashed curves denote the radii of neutrinospheres averaged over the full neutrino phase space, following the setup of Ref.~\cite{fischer2012neutrino}. 
The first shock is immediately launched after the first core bounce, resulting in enhanced electron captures on shock-dissociated protons, leading to the well studied emission of the $\nu_e$ burst of $\sim$ 5--25~ms. 
Afterwards, all neutrinospheres and the PNS start to shrink, while the shock continues to propagate outward up to a radius of $\simeq$140~km at $t_{\rm pb}\simeq 70$~ms.
Without QCD phase transition, the continuous shock retreat in spherically symmetric supernova simulations with the \texttt{s25a28} progenitor model will result in the formation of a black hole once the accumulated PNS mass exceeds the maximum mass for stable configurations~\cite{sumiyoshi2006neutrino,fischer2009neutrino,oconnor2011black}.
The infalling material falls onto the standing shock, whose radius reduces to $r\simeq 70$~km at the moment before the QCD phase transition.
Meanwhile, all neutrinosphere radii as well as the PNS radius, defined at $\rho=10^{11}$~g~cm$^{-3}$, also decrease to around 30--40~km right before the phase transition takes place. 

After the phase transition is triggered, the PNS starts to undergo an adiabatic supersonic collapse.
As a direct consequence, a second hydrodynamic shock is launched and propagates outward with a relativistic velocity of around 60\% of the speed of light.
The shock quickly reaches a radius of more than a few hundred kilometers in a few milliseconds. 
The material at the PNS surface is reheated by the shock passage and emits an intense and anisotropic flux of neutrinos of all flavors.
The neutrinosphere radii of $\nu_e$ and heavy lepton neutrinos $\nu_x$, extend to $\sim 80$~km at maximum, while the maximum value is $\sim 70$~km for $\bar\nu_e$ [see Figs.~\ref{fig:2d_thermo}(a) and (b)]. 
After $\sim$2~ms, the post-shocked material ahead of the PNS falls back again, which launches another weaker shock wave with moderately relativistic velocities on the order of few percent of the speed of light. 
At the same time, the PNS surface and the neutrinosphere radii undergo a phase of acoustic oscillations with a damped amplitude until $t_{\rm pb}\simeq$350~ms.
Most material above the PNS is ejected following the launch of these shocks. 
After $t_{\rm pb}=350$~ms, the density profile above the PNS surface steepens so that the neutrinosphere radii retreat back toward the compact PNS surface at $\sim$20~km.

The evolution of entropy per particle, $s$, and electron fraction, $Y_e$, around the second core bounce are shown in Figs.~\ref{fig:2d_thermo}(c)~and~(d), respectively.
For convenience, we use the post-second-bounce time $t_{\rm p2b}$ here as well as for the rest of the paper. 
As the second bounce shock carries a significant part of the gravitational binding energy associated with the second collapse and latent heat released from the first-order QCD phase transition, the material above the bounced core is substantially heated up due to the shock passage.
In the post shock layers, a thin shell characterized by unusually high entropy $s\gtrsim 100$, as indicated by the red color in Fig.~\ref{fig:2d_thermo}(c), with a moderately neutron-rich composition with $Y_e \simeq 0.35$.
This thin shell is followed by more neutron-rich ejecta with $Y_e\simeq 0.2$ at minimum. 
The low $Y_e$ values of these ejecta, which originally reside at the PNS interior at higher densities, are due to their extremely rapid expansion that limits neutrino exposures and prevents $Y_e$ from reaching values near weak equilibrium.
These ejecta are represented by the region with yellow-to-orange colors in Fig.~\ref{fig:2d_thermo}(d). 
In contrast to the preceding high-entropy ejecta, these subsequent ejecta have lower entropy $s\simeq 10$--20, which is still higher than their original entropy inside the PNS prior to the second collapse. 
Both components constitute the direct ejecta.

On a longer timescale of several tens of milliseconds after the second bounce, neutrinos begin to play an important role in heating the intermediate ejecta and reshaping the matter composition, partly due to their slower expansion timescale.
Neutrinos deposit energy into the ejecta and increase the entropy continuously, to values of $s\sim30-40$ for the region between radii of $\sim$30--150~km after $t_{\rm p2b}\sim 10$~ms.
After this phase, the continuous NDW outflow takes over.
At $t_{\rm p2b}\gtrsim\mathcal{O}(100)$~ms, the entropy in the NDW whose $Y_e\lesssim 0.5$ exceeds $s\simeq100$, significantly higher than in canonical SN explosion models, due to the greater compactness of the PNS~\cite{qian1996nucleosynthesis}. 

In addition to the general features discussed above, Fig.~\ref{fig:rt} shows the radial profiles of rest-mass density, $\rho$, temperature, $T$, and radial velocity, $v$, at eight different times, within a few tens of milliseconds after the second bounce, to provide a better understanding of the evolution of hydrodynamic quantities.
These plots show that the second bounce takes place at a radius of about 7~km, where a density jump from below the nuclear saturation density to $\rho\gtrsim 7\times 10^{14}$~g~cm$^{-3}$ appears at $t_{\rm p2b}=0$ [black line in Fig.~\ref{fig:rt}(a)].
The launched shock moves rapidly outward, characterized by a maximum velocity of $v\simeq 0.3$ at $t_{\rm p2b}=0.7$~ms [blue line in Fig.~\ref{fig:rt}(c)] and $v\simeq 0.55$ at $t_{\rm p2b}=1.4$~ms [yellow line in the Fig.~\ref{fig:rt}(c)].
The post-shocked regions are significantly heated, reaching $T\approx 5$~MeV at a radius below about 115~km at $t_{\rm p2b}=1.4$~ms [yellow line in Fig.~\ref{fig:rt}(b)].
At ${t_{\rm p2b}}=2.3$~ms [red lines in Figs.~\ref{fig:rt}(a) and (b)], both the density and temperature drop to lower values as the relativistic shock moves to increasingly larger radii.
Meanwhile, part of the material falls back onto the PNS surface, e.g., 30~km$\lesssim r \lesssim$55~km at $t_{\rm p2b}=2.3$~ms.
This leads to the aforementioned formation of an additional, weaker shock that moves radially outward, reaching radii of about 85~km at $t_{\rm p2b}\sim 6.9$~ms [yellow lines in the bottom panels of Fig.~\ref{fig:rt}].
The maximum velocity behind this shock is $v\simeq 0.04$ at a radius of about 70~km.
Afterwards, both the density and temperature profiles steepen above the PNS surface, which becomes increasingly compact. 
For instance, the density drops to $\rho\simeq 10^{9}~\mathrm{g~cm}^{-3}$ at a radius of around 30~km in a few tens of milliseconds after the second bounce.
This is consistent with the retreating neutrinosphere radii shown in Fig.~\ref{fig:2d_thermo}(b).

\section{Neutrino features}
\label{sec:neutrinos}
As discussed above, the phase transition significantly alters the neutron excess of the ejecta, the compactness of the PNS, and the dynamics of the inner ejecta, within $\mathcal{O}(10)$~ms.
All these effects shape the concurrent emission properties of neutrinos such as luminosities, energy spectra, and the flavor ratio among the different species.  

\begin{figure}[!hbt]
\centering
\includegraphics[width=0.98\columnwidth]{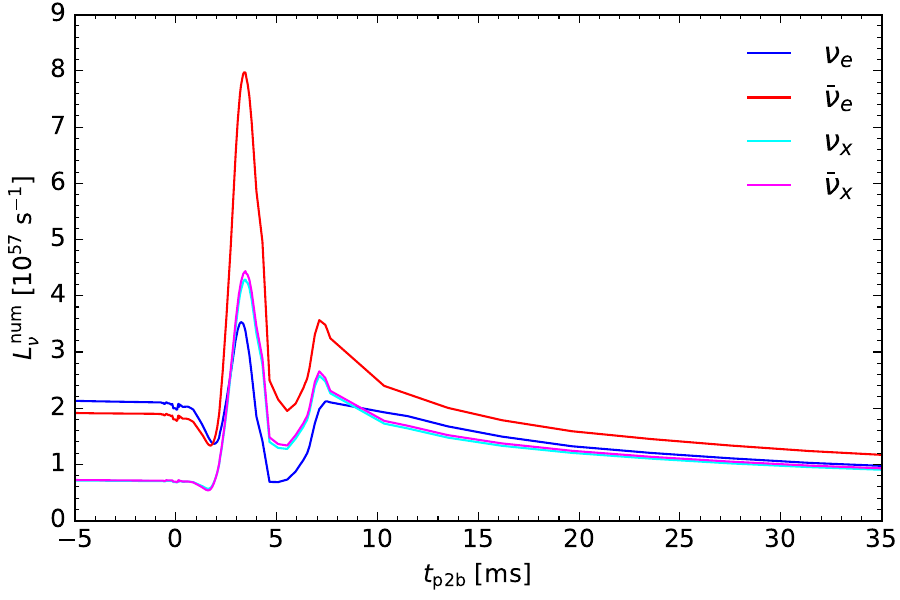}
\llap{\parbox[b]{5.6in}{\small (a)\\\rule{0ex}{1.9in}}}
\includegraphics[width=0.98\columnwidth]{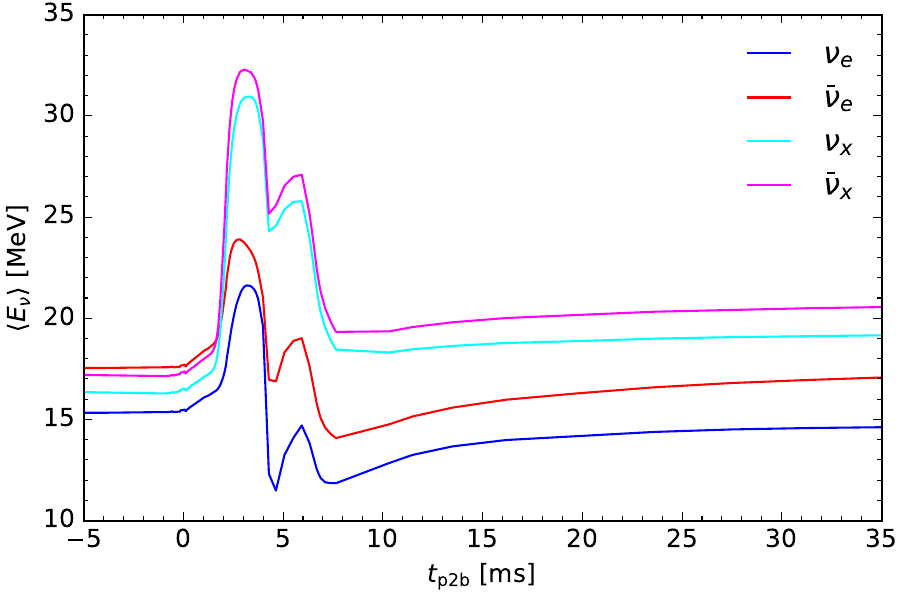}
\llap{\parbox[b]{5.6in}{\small (b)\\\rule{0ex}{1.9in}}}
\caption{\label{fig:nu_property}Evolution of the neutrino number luminosities $L_\nu^{\rm num}$ in (a) and mean energies $\langle E_\nu \rangle$ in (b) as a function of the post-second-bounce time, $t_{\rm p2b}$, sampled in the comoving frame of reference at 500~km.}
\end{figure}

\subsection{Neutrino emission properties}
Figure~\ref{fig:nu_property} shows the evolution of the neutrino number luminosity and mean energy for $-5~{\rm ms}<t_{\rm p2b}<35~{\rm ms}$. 
The neutrino number luminosity $L_\nu^{\rm num}\equiv L_\nu/\langle E_\nu \rangle$ is defined as the energy luminosity $L_\nu$ divided by the neutrino mean energy $\langle E_\nu \rangle$.
Both quantities are evaluated at a reference radius of 500~km in the comoving frame of reference, which is sufficiently far away from the PNS in the freely streaming regime.
A detailed freely streaming analysis has been conducted in Appendices~A and B of Ref.~\cite{largani2024constraining}.

Before the second bounce, the evolution of neutrino properties exhibit features similar to the canonical SNe during the accretion phase. 
The number luminosity of $\nu_e$ is larger than that of $\bar\nu_e$ by around 10\%. 
For heavy lepton flavors, the number luminosity of each neutrino species is slightly less than half of $\nu_e$ or $\bar\nu_e$ number luminosities. 
The mean energies follow the same hierarchy of $\langle E_{\nu_e} \rangle<\langle E_{\nu_x} \rangle<\langle E_{\bar\nu_e} \rangle$ as the earlier accreting phase.
Notice that the mean energies of $\nu_x$ and $\bar\nu_x$ are not the same because of the weak magnetism effect in the neutrino-nucleon scattering \cite{horowitz2002weak}.

During the PNS collapse right before the second bounce, the associated contraction of the PNS radius leads to a synchronized decrease of the neutrino number luminosities of all flavors for $\sim$1~ms, shown by the dips around $t_{\rm p2b}\simeq 1.5$~ms in Fig.~\ref{fig:nu_property}(a). 
Note that the delay time is due to the propagation of neutrinos from the neutrinospheres to the radius of 500~km. 
At the same time, the increased compactness, due to the contraction of PNS, reduces the neutrino opacity.
This allows neutrinos of higher energy to escape more easily, leading to the synchronized increase of the mean energy of all flavors.
Afterwards, the shock wave is launched and the PNS expands, which substantially enlarges and heats the radial regions responsible for neutrino production.
A prominent neutrino burst of all flavors is released, associated with an increase of their mean energy, which peaks at $t_{\rm p2b}\simeq 3-4$~ms.

While this burst enhances the neutrino number luminosities of all flavors, it is important to note that the enhancement differs for different species.
In particular, because of the shock heating, the local weak equilibrium condition of the originally very neutron-rich material ($Y_e\simeq 0.1$) above the bouncing core suddenly changes to a new condition favoring higher $Y_e$ values. 
As a result, a net protonization (or leptonization) of the material occurs, which results in the dominant emission of $\bar\nu_e$.
At the peak of the burst, around $t_{\rm p2b}=3.5$~ms, the $\bar\nu_e$ number luminosity reaches $L_{\bar\nu_e}^{\rm num}\simeq 8\times 10^{57}~\mathrm{s}^{-1}$, much larger than those of $\nu_x$ and $\bar\nu_x$ ($L_{\nu_x/\bar\nu_x}^{\rm num}\simeq 4.4\times 10^{57}~\mathrm{s}^{-1}$), as well as $\nu_e$ ($L_{\nu_e}^{\rm num}\simeq 3.5\times 10^{57}~\mathrm{s}^{-1}$).
During this period, the mean energy of heavy lepton flavors ($\langle E_{\bar\nu_x} \rangle \approx 32$~MeV) also becomes larger than $\langle E_{\bar\nu_e} \rangle \approx 23$~MeV.  
The mean energy of $\nu_e$ remains the lowest at around $\langle E_{\nu_e} \rangle \approx 21$~MeV.  

After the dominant burst related to the strong shock passage across the neutrinospheres, the neutrino mean energies of all flavors decrease momentarily to locally minimal values around $t_{\rm p2b}=4.5$~ms and arise again due to heating by the weaker shock generated from the oscillatory motion around the PNS. 
The mean energies of all flavors increase to $\langle E_{\nu_e} \rangle\simeq 14$~MeV, $\langle E_{\bar\nu_e} \rangle\simeq 19$~MeV, and $\langle E_{\nu_x} \rangle \simeq 27$~MeV, at $t_{\rm p2b}\sim 6$~ms.
Associated with this is an increase in the neutrino number luminosities of all flavors, peaking approximately 1.5~ms later.

Once the PNS settles into a quasi-stationary state that gradually cools off, all neutrino number luminosities start to decrease continuously. 
The decreasing rate of the $\nu_e$ number luminosity is slower than those of heavy lepton flavors, so that the hierarchy between them switches at $t_{\rm p2b}\simeq 9$~ms.
The order of neutrino mean energies among different species does not change during this phase, while their values slightly increase for several tens of milliseconds as it becomes easier for neutrinos with higher energies to decouple from matter, before the eventual PNS cooling sets in. 

\begin{figure}[!hbt]
\centering
\includegraphics[width=0.98\columnwidth]{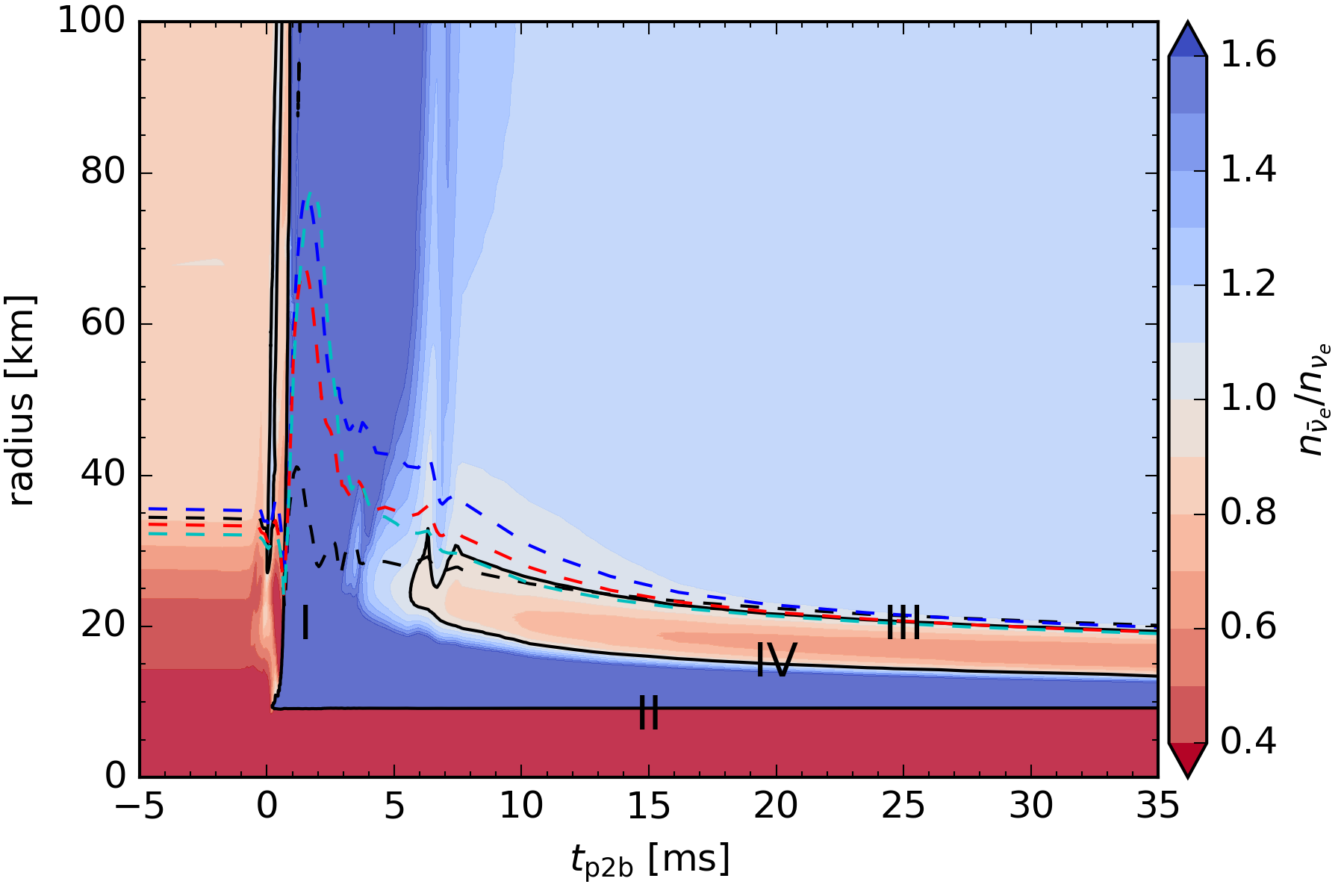}
\llap{\parbox[b]{5.6in}{\small (a)\\\rule{0ex}{1.9in}}}
\includegraphics[width=0.98\columnwidth]{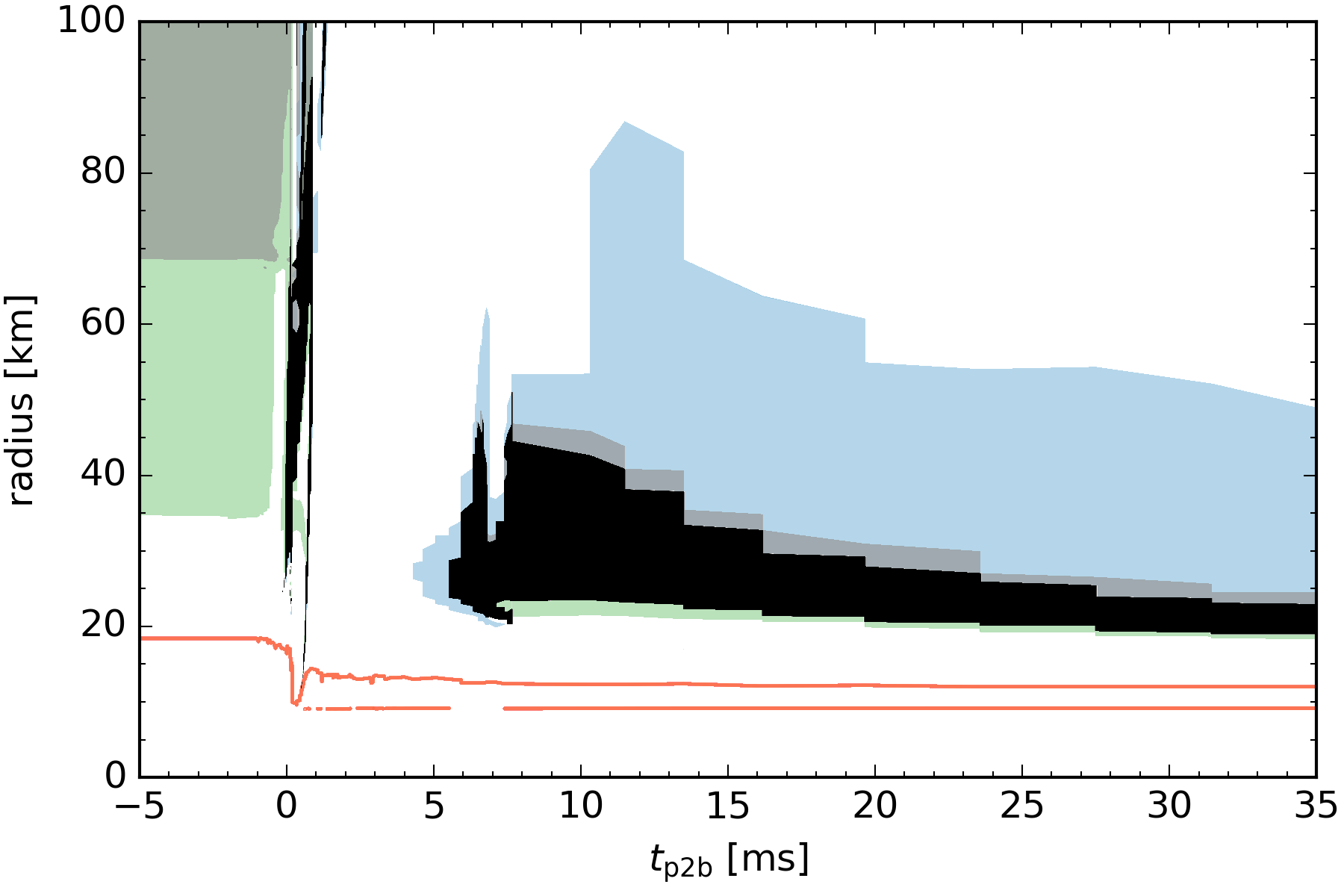}
\llap{\parbox[b]{5.6in}{\small (b)\\\rule{0ex}{1.9in}}}
\caption{\label{fig:2d_region}Evolution of radial profiles of the ratio of neutrino number densities $n_{\bar\nu_e}/n_{\nu_e}$ (a) and regions associated with FFI (b) as functions of $t_{\rm p2b}$.
Black, blue, red, and cyan dashed curves (a) show the radii of PNS, $\nu_e$-, $\bar\nu_e$-, and $\nu_x$-spheres, respectively. Black solid curves mark four contour lines of $n_{\bar\nu_e}/n_{\nu_e}=1$.
Both black and gray domains (b) show the existence of ELN angular crossing.
Considering more variations may be achieved in multi-dimensional SN simulations, the green and blue domains show where FFI exists when the angular distribution of $\bar\nu_e$ is amplified by 12\% or suppressed by 10\%, respectively.
Below the pink curves (b) is the region prohibiting any collective flavor instability due to the violation of $\nu$ELN.}
\end{figure}

\subsection{Local neutrino-antineutrino asymmetry}
The evolution of the basic neutrino emission properties around the second bounce discussed above are tied to the presence of FFI near the neutrino decoupling region.
As the existence of the local energy-integrated $\nu$ELN angular crossing is a sufficient and necessary condition for FFI~\cite{morinaga2022fast}, this implies that FFI should be present in regions where the asymmetry factor, defined as the ratio of $\nu_e$ and $\bar\nu_e$ number densities, becomes 1.  
This is because when $n_{\bar\nu_e}/n_{\nu_e}=1$, any slight difference in the $\nu_e$ and $\bar\nu_e$ angular distributions guarantees a $\nu$ELN angular crossing, assuming that $\bar\nu_x$ and $\nu_x$ share exactly the same angular distribution. 
On the other hand, when $n_{\bar\nu_e}/n_{\nu_e}\gg 1$ or when $n_{\bar\nu_e}/n_{\nu_e}\ll 1$, it requires the two species to have very different angular distributions to obtain an angular crossing.

Figure~\ref{fig:2d_region}(a) shows the radial profiles of the ratio $n_{\bar\nu_e}/n_{\nu_e}$ as a function of $t_{\rm p2b}$, with solid black contours denoting the locations corresponding to $n_{\bar\nu_e}/n_{\nu_e}=1$. 
The effect of the protonization on the ratio $n_{\bar\nu_e}/n_{\nu_e}$ due to the heating of the phase transition shock is evident. 
After the second bounce, almost the entire region above the bounced core at a radius of about 9~km has $n_{\bar\nu_e}/n_{\nu_e}>1$, except a shell with a width of 5--10~km right below the PNS surface, where $0.6\lesssim n_{\bar\nu_e}/n_{\nu_e}\lesssim 1$, after $t_{\rm p2b}\simeq 5.5$~ms.
The formation of this shell is related to the relatively faster cooling compared to the inner parts, which, in turn, leads to a local thermal equilibrium state of higher electron degeneracy, favoring lower values of $n_{\bar\nu_e}/n_{\nu_e}$. 
Note that this region corresponds to the $Y_e\sim 0.15$ shell just below the PNS surface in Fig.~\ref{fig:2d_thermo}(d) shown by red color, indicating that deleptonization takes place locally inside the PNS. 
As a result, there are two major contours of $n_{\bar\nu_e}/n_{\nu_e}=1$.
The first one separates the post-second-shock region above the inner core, labeled by I and II in Fig.~\ref{fig:2d_region}(a). 
The second one encloses the aforementioned shell with $n_{\bar\nu_e}/n_{\nu_e}<1$ below the PNS, denoted by labels III and IV.
Note that there is another contour curve of $n_{\bar\nu_e}/n_{\nu_e}=1$ at radii greater than about 27~km, that only exists very briefly for about 0.4~ms up to 100~km. 
This is related to the change of local thermodynamic conditions during the final stage of collapse that leads to the reduction of the electron degeneracy, hence the reduced equilibrium $\nu_e$ chemical potential.

Figure~\ref{fig:2d_region}(a) also shows that several regions contain strong $\nu_e$--$\bar\nu_e$ asymmetry that should prevent the existence of energy-integrated ELN angular crossings. 
The highly-degenerate PNS interior corresponding to radii less than about 24~km before the second bounce and beneath the contour line II after the bounce is dominated by electron neutrinos with $n_{\bar\nu_e}/n_{\nu_e}<0.5$. 
The efficient protonization and the associated dominant emission of $\bar\nu_e$ lead to $n_{\bar\nu_e}/n_{\nu_e}\gtrsim 1.6$ above the bounced core for $0\lesssim t_{\rm p2b}\lesssim 5$~ms as well as the radial range between the contours labeled by II and IV.
As a result, no FFI is expected in these domains.

\subsection{Regions of FFI}
While the above analysis of the asymmetry factor provides insight on the location of FFI, the discrete-ordinate neutrino transport adopted by the \textsc{boltztran} module provides detailed angular distributions of all flavors, which allows a direct identification of ELN angular crossings.
Both black and gray shaded areas in Fig.~\ref{fig:2d_region}(b) are regions where angular crossings exist based on the simulation data. 
In particular, the gray regions correspond to those with ``shallow'' angular crossing, where the crossing would have disappeared if the local angular distribution of $\nu$ELN density varies by $\sim 10^{30}~{\rm cm}^{-3}$ (around $0.1\%$ of neutrino number density near the neutrinospheres).
The shallow crossings primarily reside in the \textit{canonical} SN accreting phase before the second bounce at large radii greater than about 70~km.  
These crossings originate from the backward neutrino-nucleus scattering~\cite{morinaga2020fast} and are not expected to result in a significant amount of flavor conversion~\cite{abbar2022suppression}.

For the non-shallow crossings, they mainly show up in two different domains.
The first domain lies around the $n_{\bar\nu_e}/n_{\nu_e}=1$ contour segment I. 
They are associated with the change of matter composition during the final stage of collapse as well as the passage of the shock.
These crossings are confined within a time window of less than 1~ms, corresponding to mass shells that move rapidly outward together with the shock wave. 
Subsequently, the strong $\bar\nu_e$-dominated condition, due to protonization above the bounced core, inhibits the occurrence of FFI, until $n_{\bar\nu_e}/n_{\nu_e}$ reduces at $t_{\rm p2b}\simeq 5.5$~ms.
In particular, the second domain of FFIs coincides with the shell enclosed by the contour segments III and IV below the PNS surface discussed in the previous subsection.

In principle, FFIs should exist around the contour segment II due to the transition from small $n_{\bar\nu_e}/n_{\nu_e}$ to large values.
However, as this region lies deeply inside the PNS and neutrinospheres, the neutrino distribution is approximately isotropic. 
As a result, FFI is limited within a rather thin shell where the asymmetry between $\nu_e$ and $\bar\nu_e$ is small, hardly resolved by resolution of the hydrodynamic simulations. 
Moreover, this implies the $\nu_e$ and $\nu_x$ equilibrium chemical potentials in this thin shell are both nearly zero, indicating that the impact of any flavor conversion is negligible \cite{xiong2024fast}.

Besides the two domains discussed above, FFIs may also exist in extended neighboring regions due to the multidimensional effects such as turbulence and convection that cannot be modeled by the spherically symmetric SN simulation, which may affect the $n_{\bar\nu_e}/n_{\nu_e}$ asymmetry~\cite{tamborra2014selfsustained,nagakura2020systematic}.
Assuming that multidimensional effects can lead to changes of $n_{\bar\nu_e}$ by a factor between $-10\%$ and $12\%$ without affecting the normalized shape of its angular distribution, it substantially enlarges the regions with FFIs as shown by the green and blue domains in Fig.~\ref{fig:2d_region}(b). 
Because $\nu_e$'s dominate over $\bar\nu_e$'s during the \textit{canonical} SN accreting phase, an increase of $12\%$ for $n_{\bar\nu_e}$ (equivalently to a $\sim 10\%$ decrease of $n_{\nu_e}$) extends the FFI region down to $r\sim 35$~km, even reaching inside the neutrinospheres of both $\nu_e$ and $\bar\nu_e$.
In contrast, when a $10\%$ decrease is applied to $n_{\bar\nu_e}$, the unstable region is extended to larger radii of around 50--70~km in general, reaching a maximum radius $\sim 85$~km at $t_{\rm p2b}\sim 11$~ms. 

In addition, other types of flavor instabilities, including the slow and collisional ones (see, e.g., \cite{shalgar2023neutrino1,johns2023collisional,johns2022collisional,xiong2023evolution,xiong2023collisional,kato2023flavor,akaho2024collisional,liu2023universality}) can exist.
The pink curves in Fig.~\ref{fig:2d_region}(b) delineates the boundary below which no collective flavor instabilities should take place due to the constraint from the conservation of $\nu$ELN \cite{dasgupta2022collective}. 
Below the pink curves, no crossing exists in the full neutrino energy-angle correlated distributions, which forbids any collective flavor instabilities. 
The only exception is the horizontal line at $r\sim 9$~km corresponding to the contour segment II shown in Fig.~\ref{fig:2d_region}(a). 
Although slow and collisional flavor instabilities may exist in the domain above the pink curve at $r \sim 13$~km, we defer a thorough investigation on these instabilities to future work. 

\begin{figure*}[!hbt]
\centering
\begin{minipage}[t]{\columnwidth}
\includegraphics[width=\columnwidth]{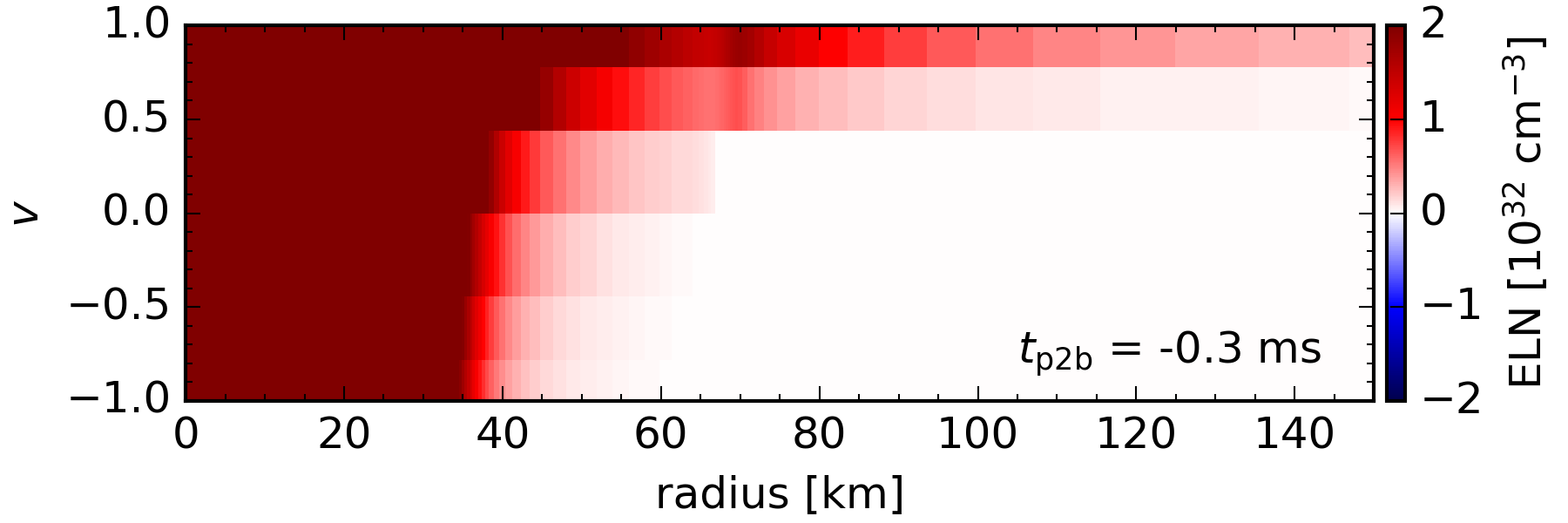}
\includegraphics[width=\columnwidth]{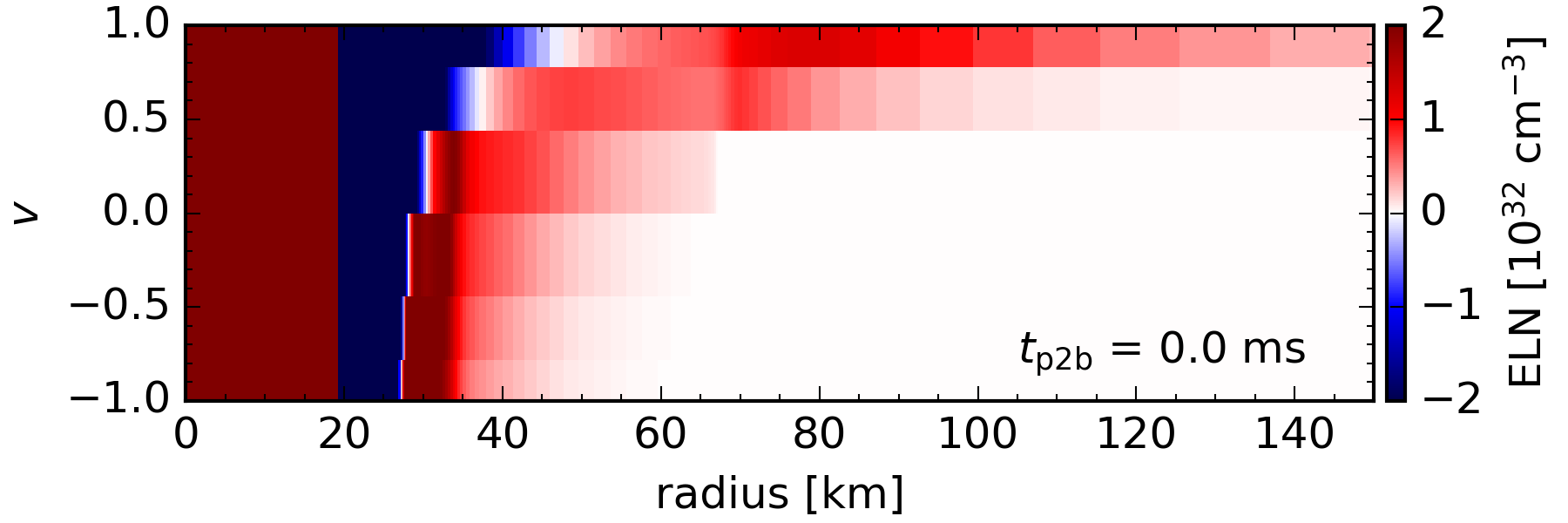}
\includegraphics[width=\columnwidth]{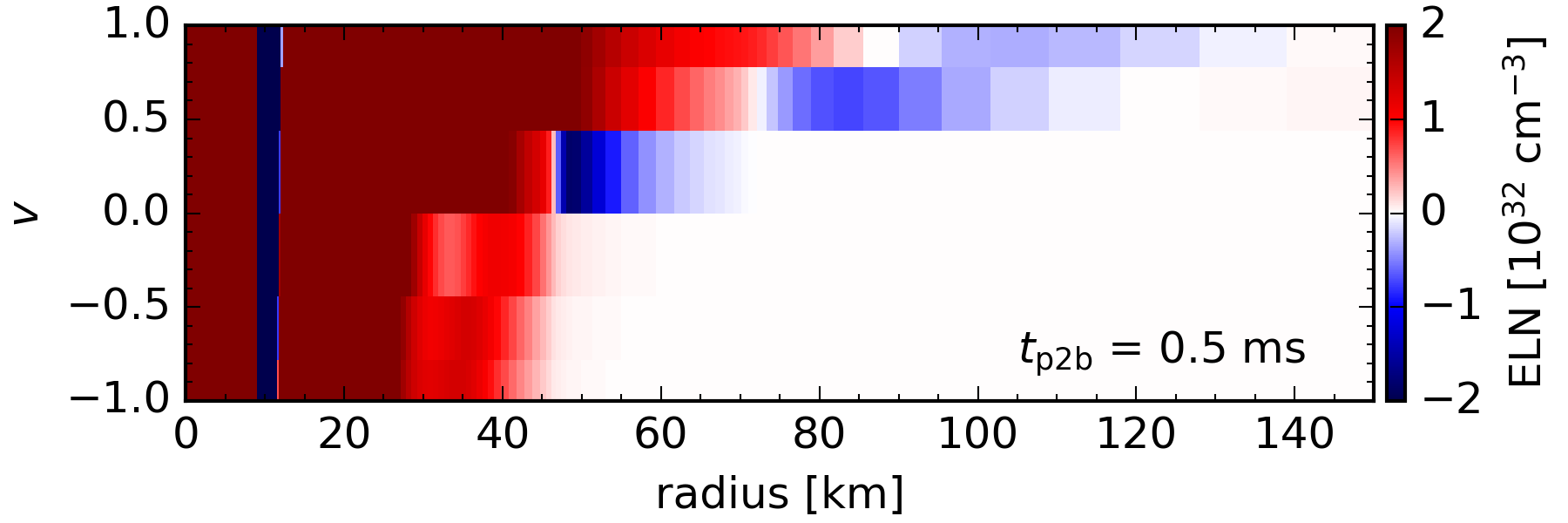}
\includegraphics[width=\columnwidth]{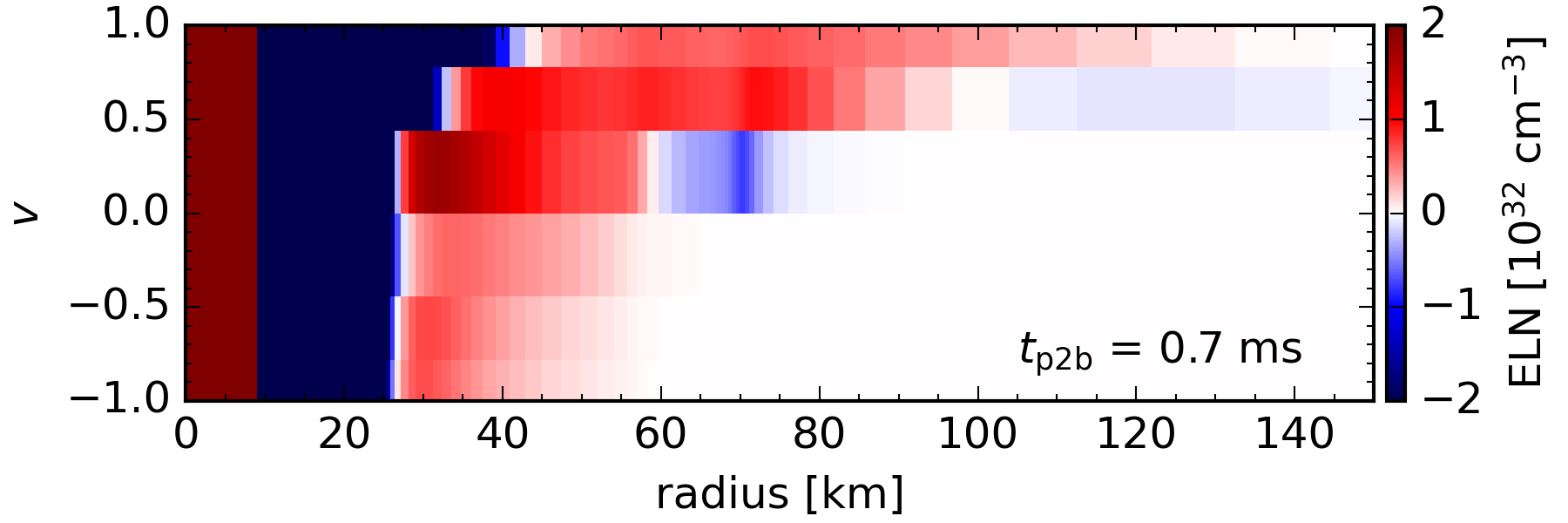}
\includegraphics[width=\columnwidth]{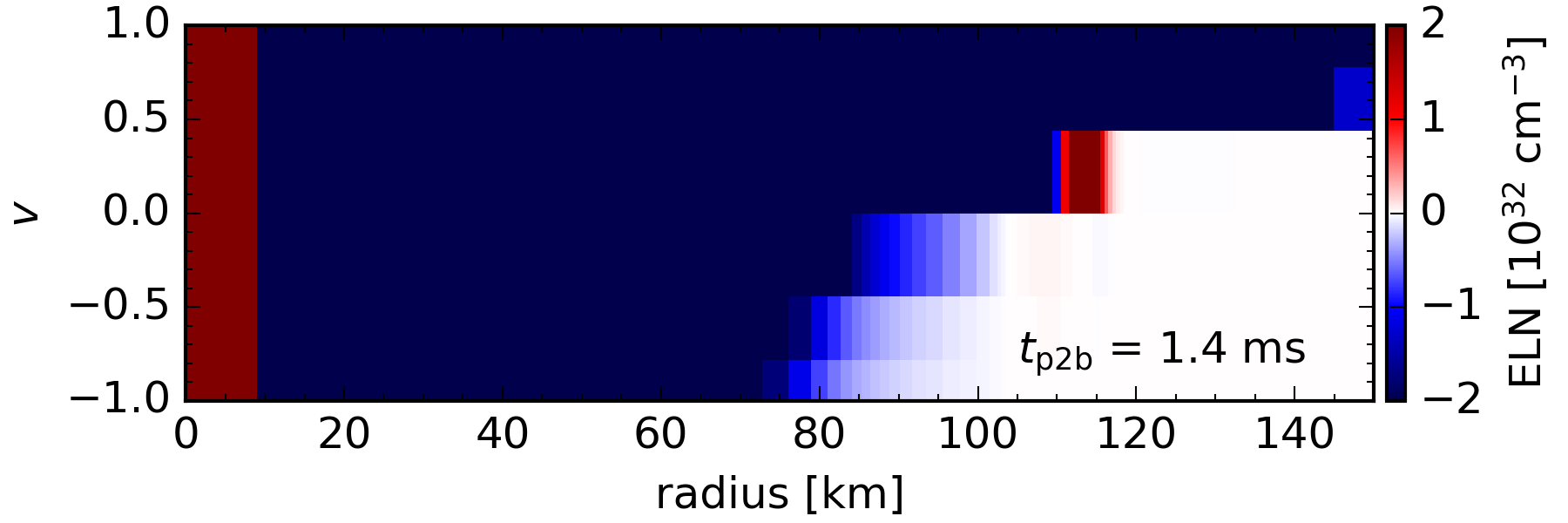}
\end{minipage}
\begin{minipage}[t]{\columnwidth}
\includegraphics[width=\columnwidth]{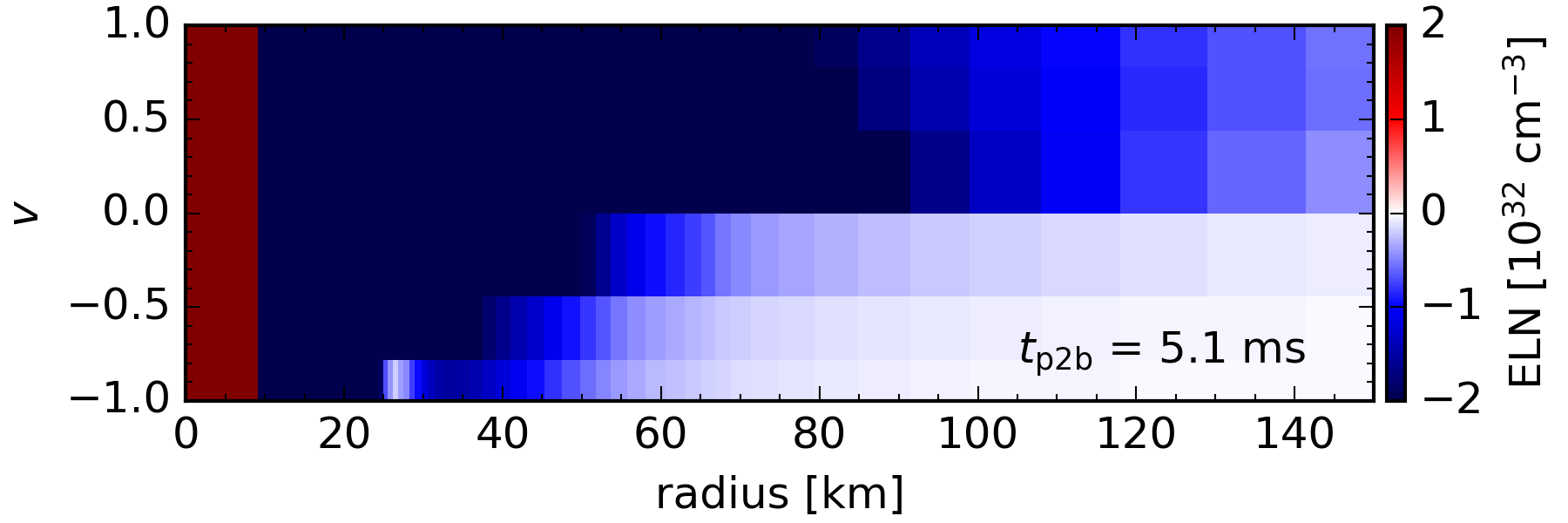}
\includegraphics[width=\columnwidth]{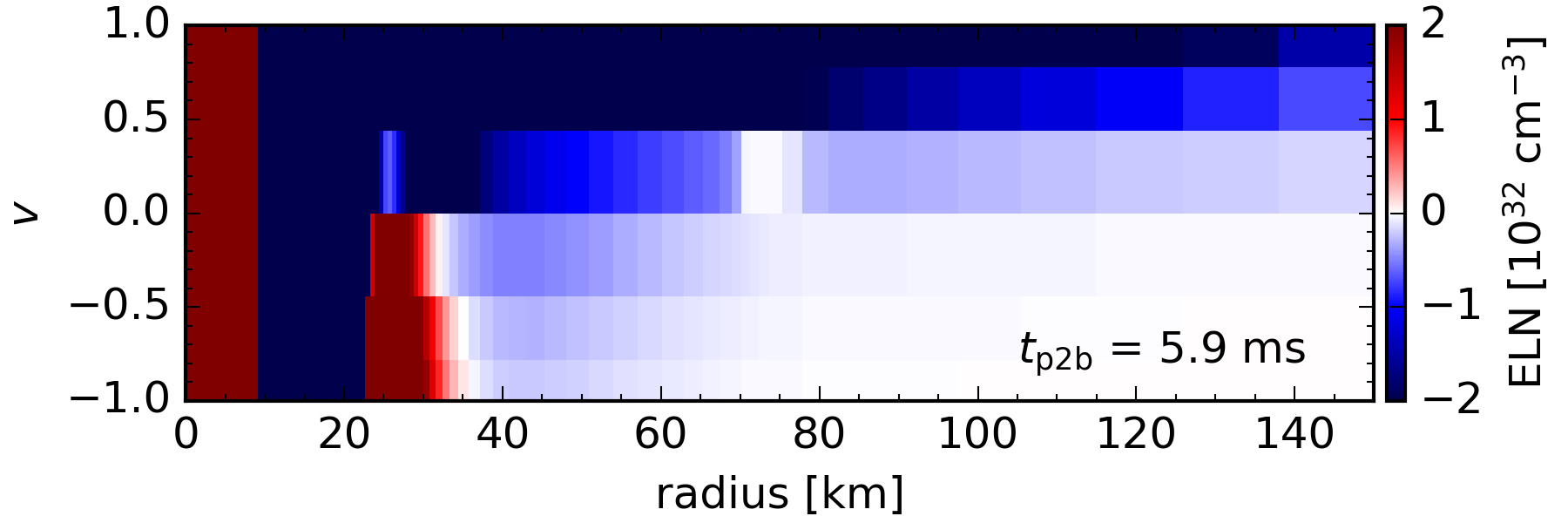}
\includegraphics[width=\columnwidth]{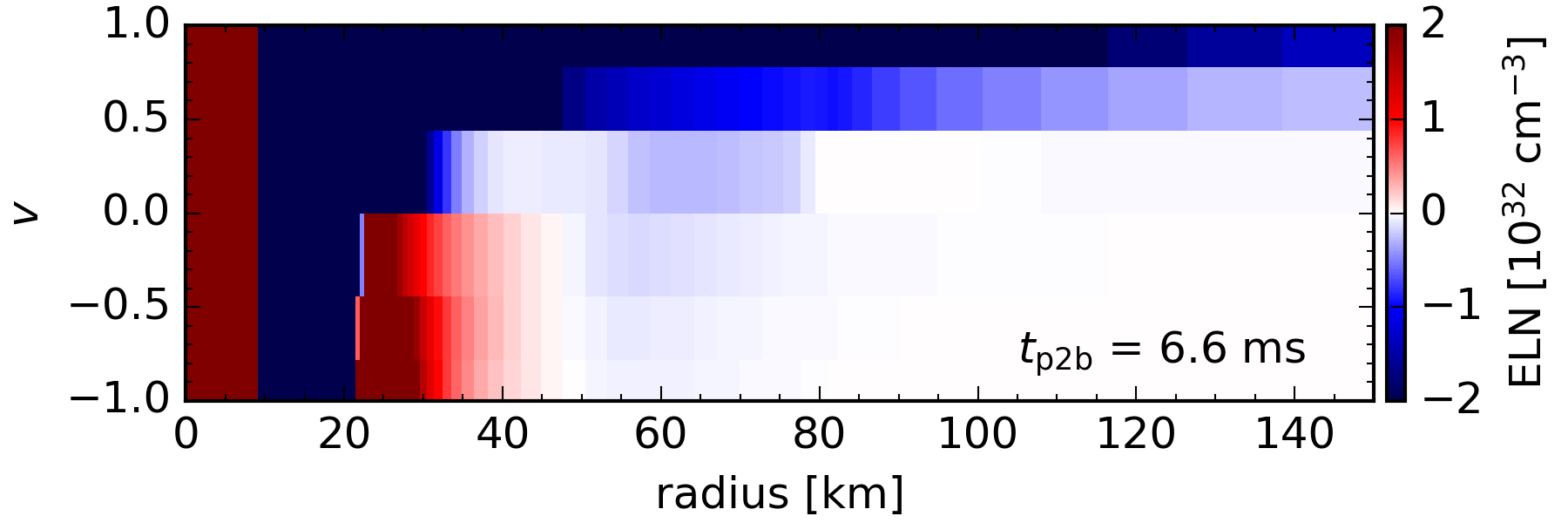}
\includegraphics[width=\columnwidth]{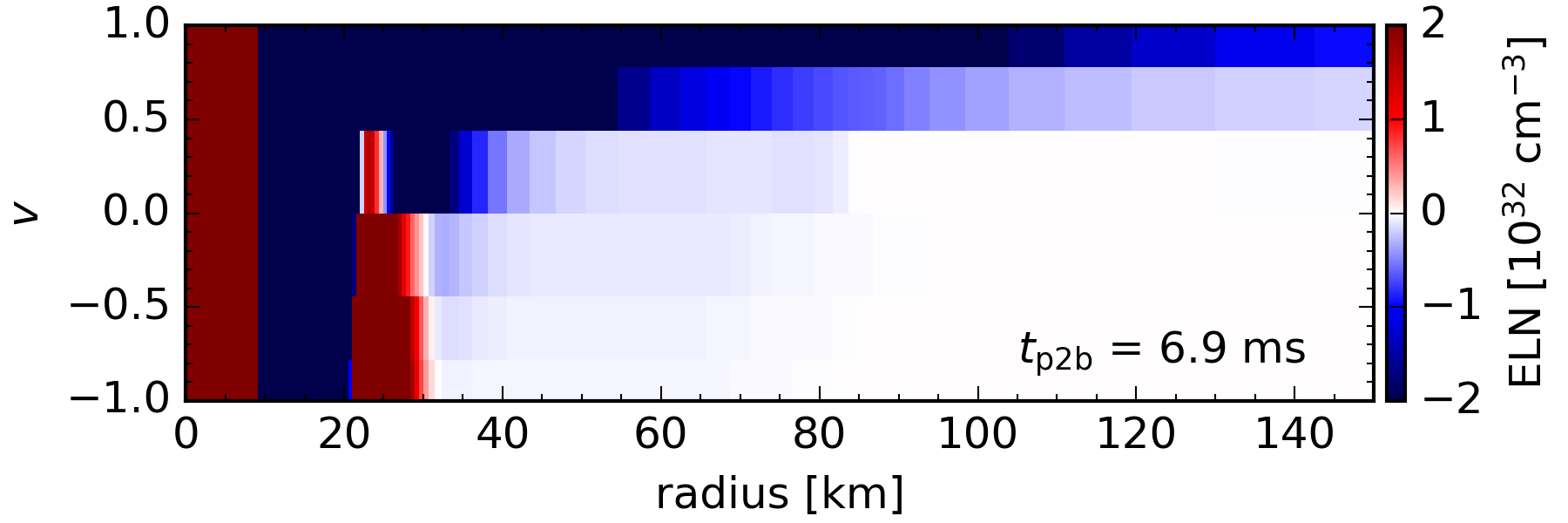}
\includegraphics[width=\columnwidth]{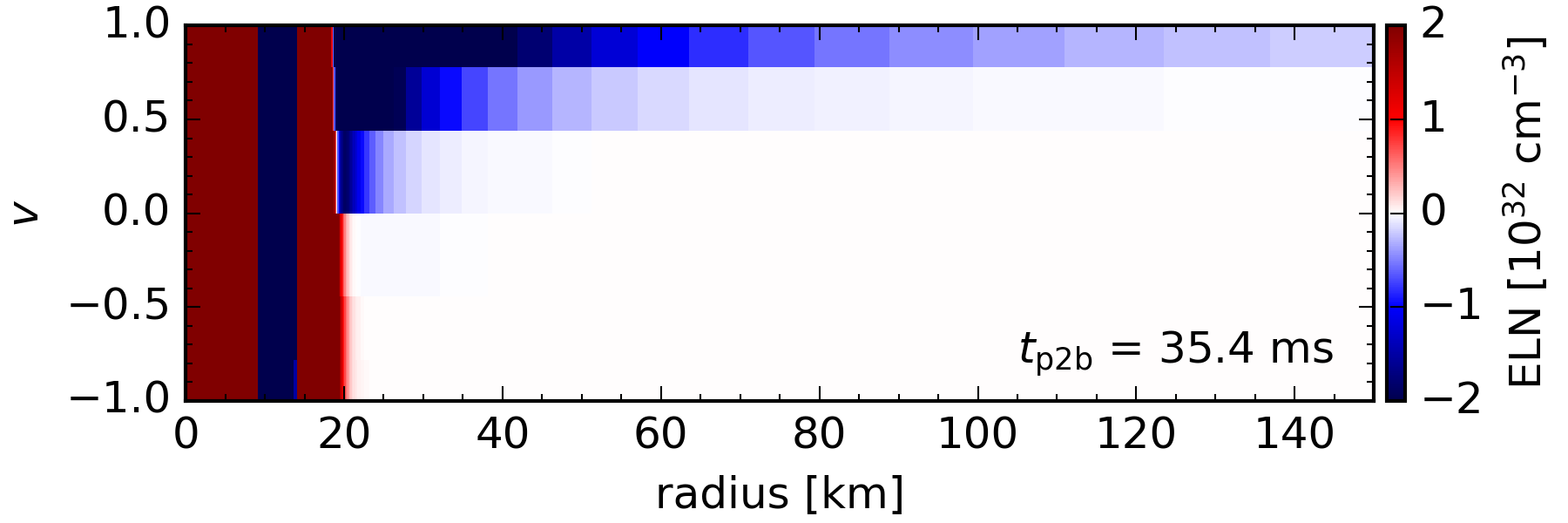}
\end{minipage}
\caption{Neutrino ELN angular distributions showing radial profiles of the radial velocity $v$, for different times $t_{\rm p2b}$.}
\label{fig:ELN}
\end{figure*}

\subsection{ELN angular distributions}
Figure~\ref{fig:ELN} shows the energy-integrated ELN angular distribution for ten snapshots to further elaborate on the emergence of the ELN crossings and their relations with the phase-transition dynamics.
The panels shown in the left (right) column correspond to snapshots taken during the first (second) phase of the FFIs aforementioned. 

At bounce ($t_{\rm p2b}=0$), the ELN in regions between $20~\mathrm{km}~\lesssim~r~\lesssim~30~\mathrm{km}$ becomes negative.
The angular crossings exist at $30~\mathrm{km}~\lesssim~r~\lesssim~45$~km due to the reason discussed in the previous section.
The crossing velocity appears around $v\gtrsim -1$ at a radius of about 30~km and moves to $v\lesssim 1$ at a radius of about 45~km.
After bounce, the ELNs in regions swiped by the strong shock wave become dominated by $\bar\nu_e$ as shown by the blue horizontal band between 9~km~$\lesssim~r~\lesssim~12$~km at $t_{\rm p2b}=0.5$~ms and 8~km~$\lesssim~r~\lesssim~26$~km at $t_{\rm p2b}=0.7$~ms.
ELN crossings appear right above the shock and move rapidly outward.

Afterwards, the second phase of FFI emerges at $t=5.9$~ms above a radius of about 22~km due to the cooling of the outer part of the PNS, which re-increases the $\nu_e$ degeneracy as discussed before.
As shown in the three middle panels at the right column of Fig.~\ref{fig:ELN}, the transition from $n_{\bar\nu_e}/n_{\nu_e}>1$ to $n_{\bar\nu_e}/n_{\nu_e}<1$ below the neutrinospheres, together with the fact that the $\bar\nu_e$ angular distribution is more forward-peaked than that of $\nu_e$ due to the earlier decoupling, lead to the formation of regions where $\nu_e$ ($\bar\nu_e$) dominates the backward (forward) propagating direction above $n_{\bar\nu_e}/n_{\nu_e}\simeq 1$. 
The PNS cools off but continuously emits more $\bar\nu_e$'s than $\nu_e$'s as the consequence of protonization
The same feature remains until $t_{\rm p2b}=35.4$~ms shown in the lower right panel of Fig.~\ref{fig:ELN}, indicating the robust presence of ELN crossings in QCD phase transition SNe.

\subsection{Progenitor dependence}
\label{subsec:progenitor}

The occurrence times of the QCD phase transition and second shock bounce also depend on the progenitor properties.
To illustrate the difference and similarity in a different progenitor mass, we repeat the same analysis for the model {\tt s40a28} RDF-1.2 from \cite{largani2024constraining}.
This SN model starts from a stellar progenitor star of $40~M_\odot$ with solar metallicity from Ref.~\cite{rauscher2002nucleosynthesis}, with the second shock occurring at a longer post-bounce time $t_{\rm pb}\sim 923$~ms.

\begin{figure}[!hbt]
\centering
\includegraphics[width=0.98\columnwidth]{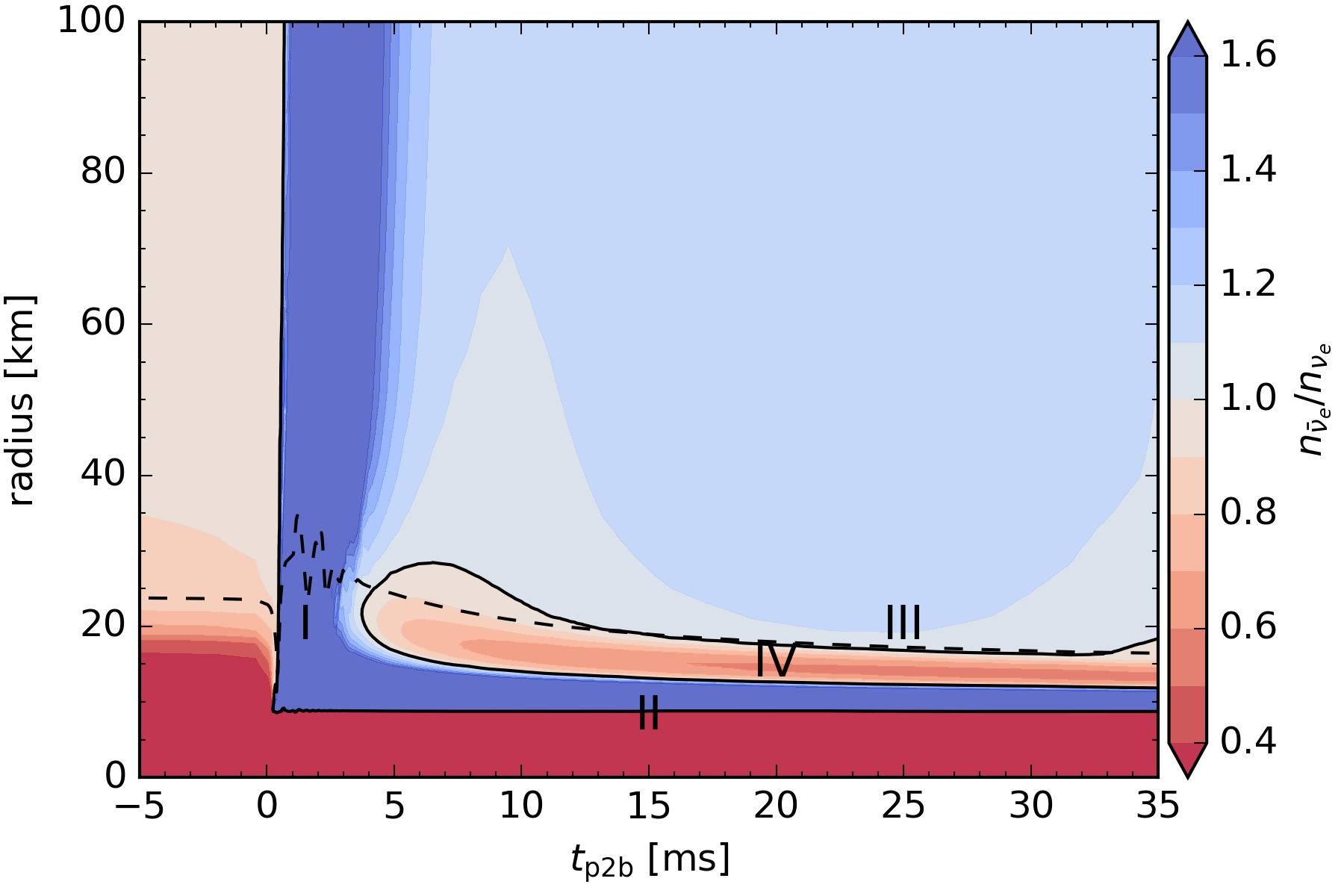}
\llap{\parbox[b]{5.6in}{\small (a)\\\rule{0ex}{1.9in}}}
\includegraphics[width=0.98\columnwidth]{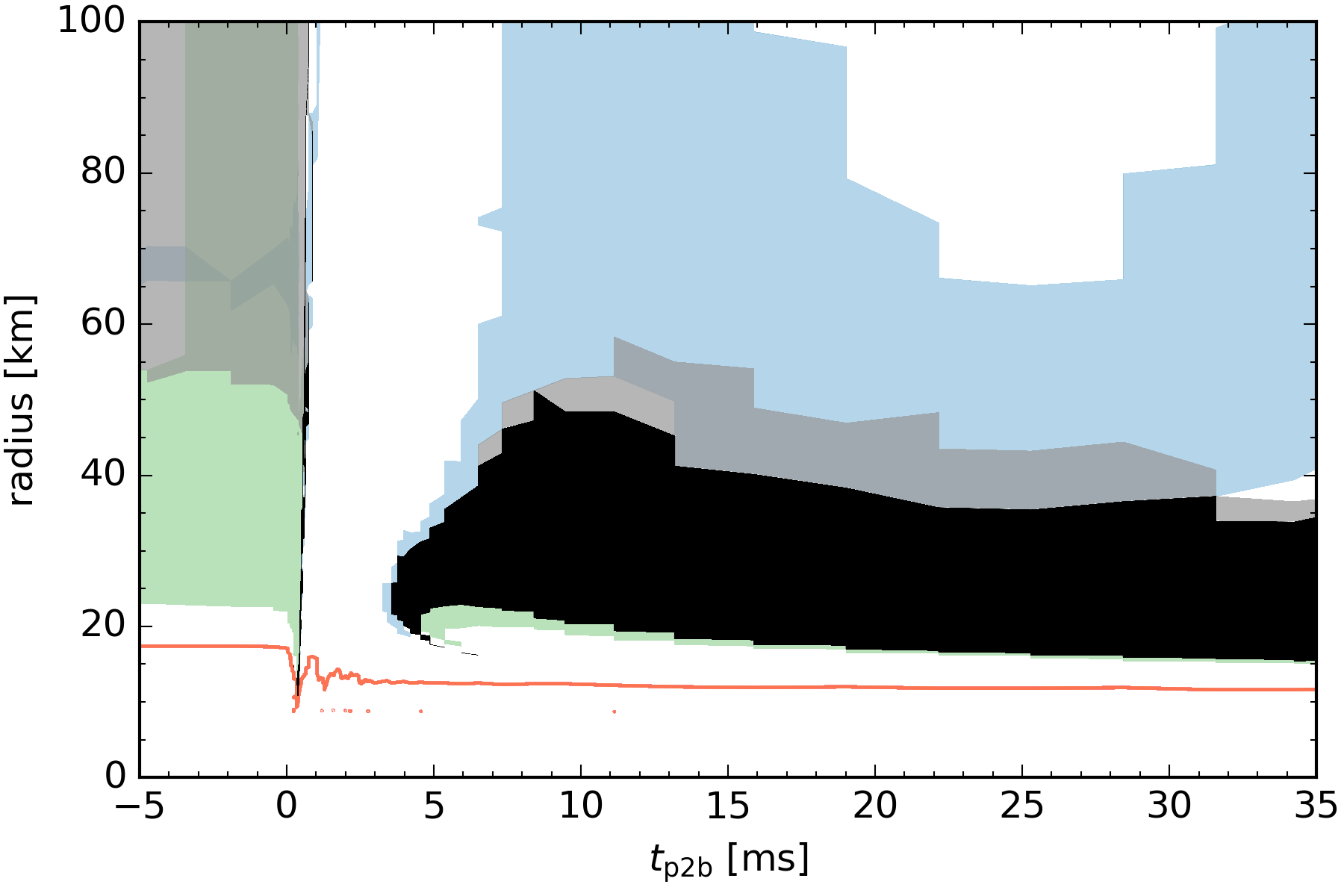}
\llap{\parbox[b]{5.6in}{\small (b)\\\rule{0ex}{1.9in}}}
\caption{\label{fig:2d_region_s40a28}Evolution of radial profiles of the ratio of neutrino number densities $n_{\bar\nu_e}/n_{\nu_e}$ (a) and regions associated with FFI (b) as functions of $t_{\rm p2b}$ for the SN model {\tt s40a28} RDF-1.2.
Definitions of legends and regions are same as in Fig.~\ref{fig:2d_region}.}
\end{figure}

Figure~\ref{fig:2d_region_s40a28}(a) shows the radial profiles of the ratio $n_{\bar\nu_e}/n_{\nu_e}$ in the same fashion as in Fig.~\ref{fig:2d_region}.
Before the second bounce, the PNS radius is about 22~km, slightly less than the value in the {\tt s25a28} model.
The rapid protonization accompanied with the shock heating leads to a prompt switch from the condition of $n_{\bar\nu_e}/n_{\nu_e}\lesssim 1$ to $n_{\bar\nu_e}/n_{\nu_e} > 1$ above the inner core at $t_{\rm p2b}\sim 0$, which is separated by the contour lines I and II.
Correspondingly, a fast moving region for FFI is spotted in Fig.~\ref{fig:2d_region_s40a28}(b) at $t_{\rm p2b}\sim 0$--1~ms.

After $\sim 4$~ms, the neutrino emission near the top of the PNS starts to favor the electron neutrino instead of antineutrino.
A region of $n_{\bar\nu_e}/n_{\nu_e}<1$ surrounded by the contours III and IV emerges due to the cooling, which overlaps with the second phase of FFI, ranging from $\sim 20$~km to $\sim 50$~km at most at $t_{\rm p2b}\sim 10$~ms.
When considering the potential variation of the $\bar\nu_e$ flux due to multidimensional effects, the green and blue bands show the extended regions with FFI during the canonical SN evolution and after the phase transition, respectively.
Neither of them results in new FFI in the intermediate region from $t_{\rm p2b}\sim 1$~ms to 4~ms due to the large amount of asymmetry between $\bar\nu_e$ and $\nu_e$.
These results suggest that the mechanism resulting in the presence of FFCs in QCD SN is general and qualitatively unaffected by the progenitor masses.

\section{Investigations of fast flavor instability}
\label{sec:FFI}
In this section, we investigate the properties of FFI mainly based on the SN model {\tt s25a28} RDF-1.9 and discuss its potential impact on the neutrino signal and nucleosynthesis.

\begin{figure}[!hbt]
\centering
\includegraphics[width=0.98\columnwidth]{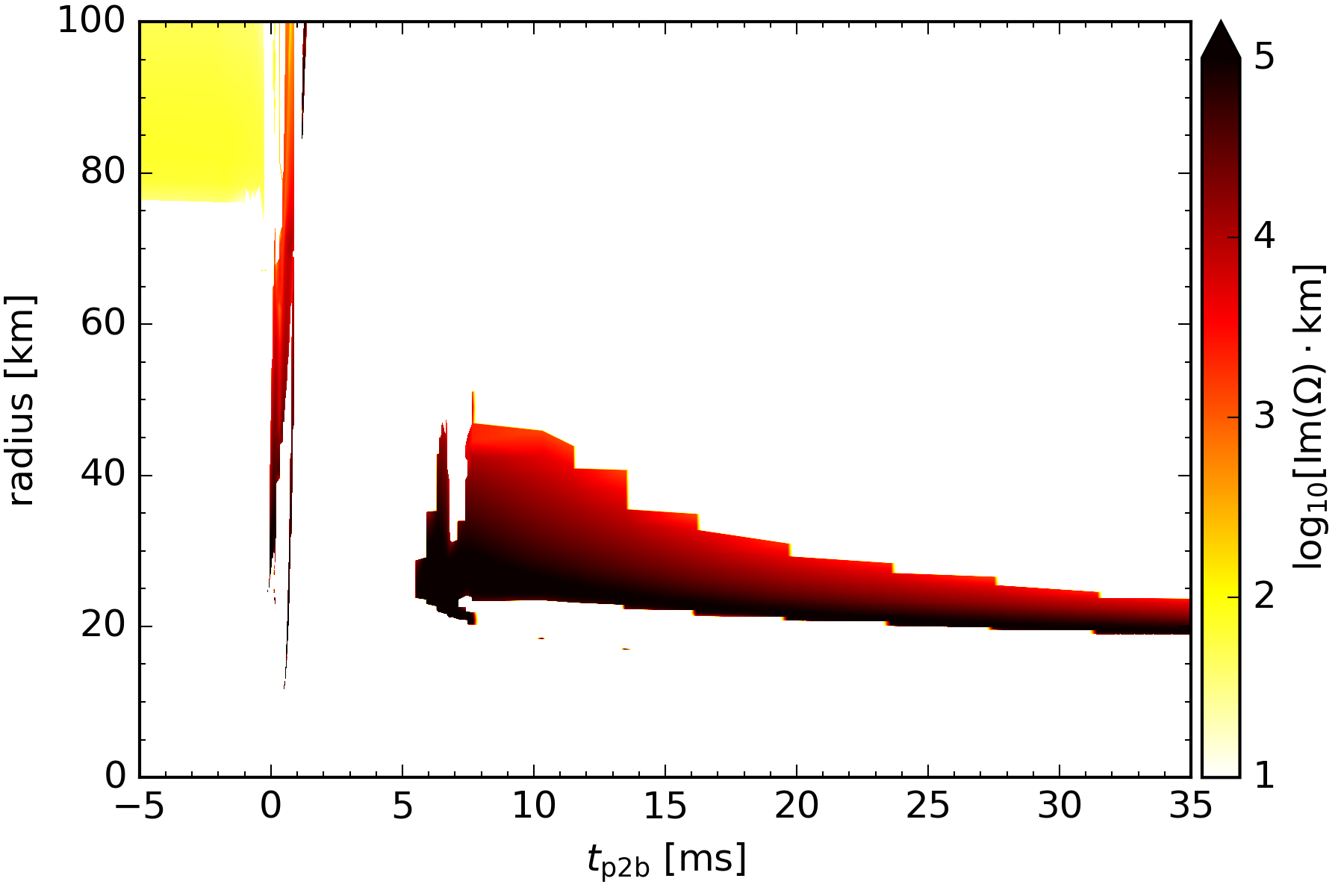}
\llap{\parbox[b]{5.6in}{\small (a)\\\rule{0ex}{1.9in}}}
\includegraphics[width=0.98\columnwidth]{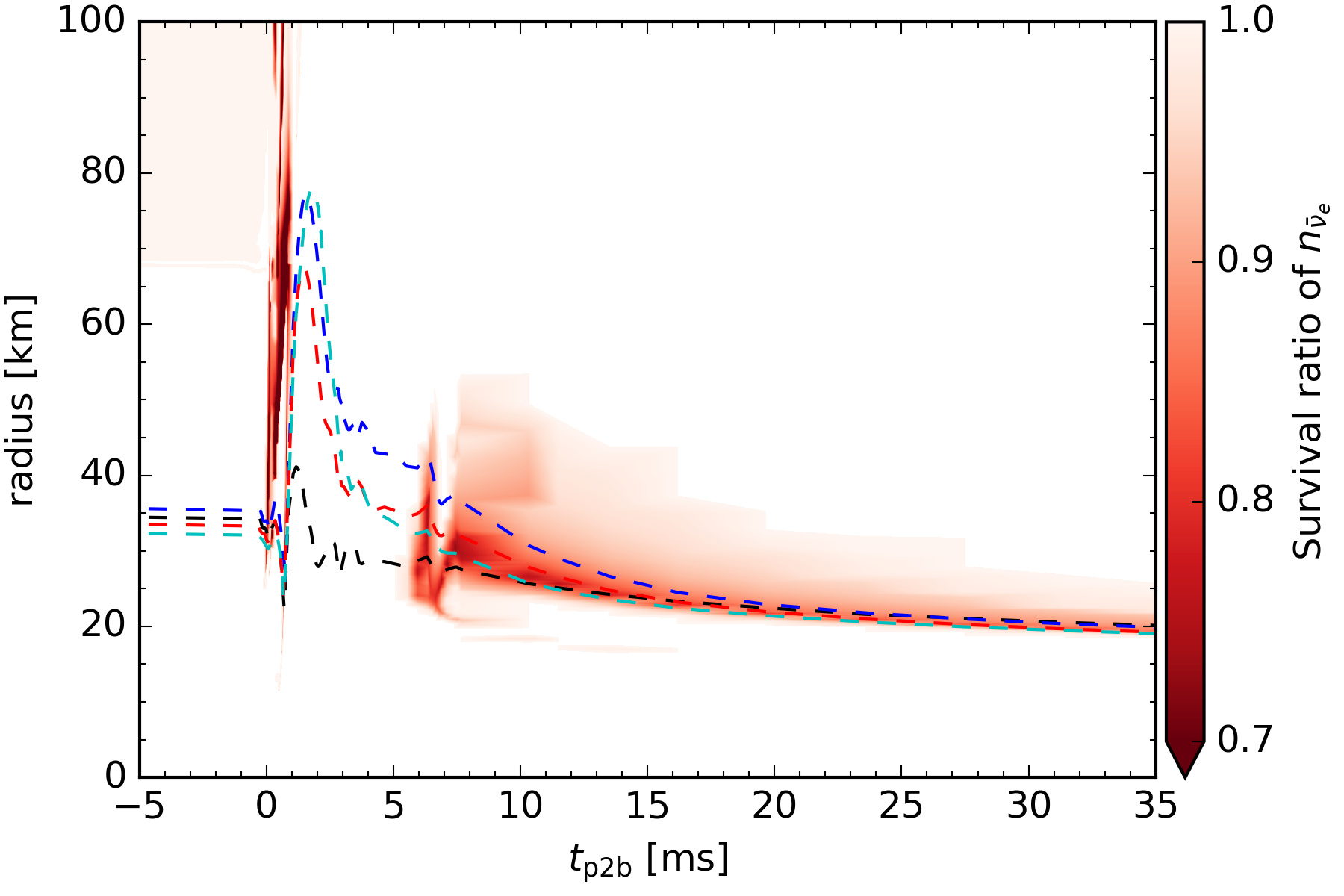}
\llap{\parbox[b]{5.6in}{\small (b)\\\rule{0ex}{1.9in}}}
\includegraphics[width=0.98\columnwidth]{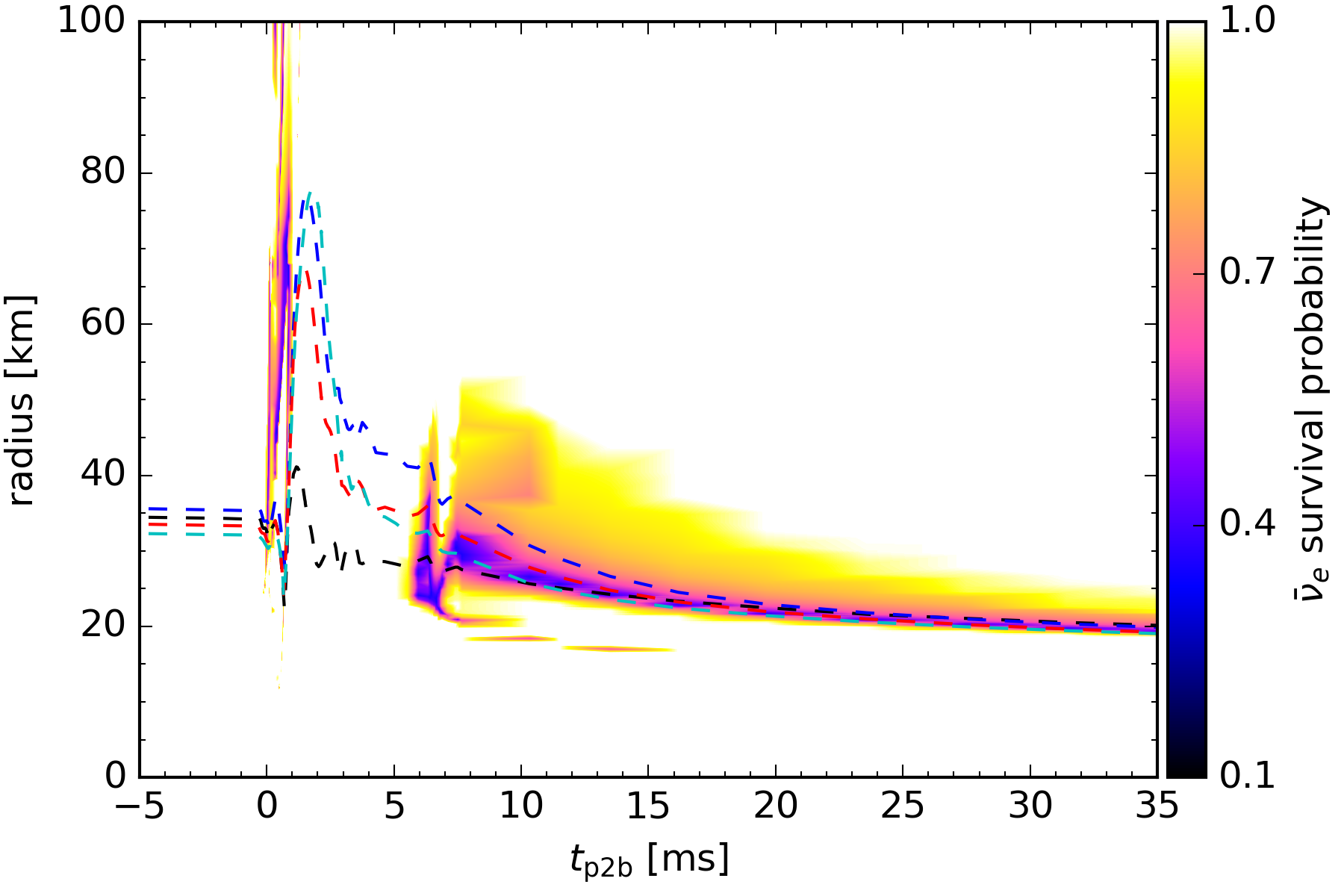}
\llap{\parbox[b]{5.6in}{\small (c)\\\rule{0ex}{1.9in}}}
\caption{\label{fig:2d_ImOmega} Maximum growth rate $\mathrm{Im}(\Omega)$ (a), $n_{\bar\nu_e}$ survival ratio (b), and angle-integrated $n_{\bar\nu_e}$ survival probability (c) as a function of time $t_{\rm p2b}$ and radius $r$. Black, blue, green, and cyan dashed curves in (b) and (c) show the radii of PNS, $\nu_e$-, $\bar\nu_e$-, and $\nu_x$-spheres, respectively.}
\end{figure}

\subsection{Linear stability analysis and FFI growth rates}
We perform the two-flavor linear stability analysis (LSA) between the electron and muon flavors for the FFI, following the approach used in, e.g., \cite{xiong2023evolution,xiong2024fast}. 
Note that we neglect the potential impact of other type of instabilities, the angular advection, as well as effects due to inhomogeneous matter~\cite{bhattacharyya2025role} and neutrino number densities.

For each time $t$, the linearized equation in a local region surrounding radius $r$ is
\begin{align}
    ( \partial_t + v \partial_r) \varrho & = -i[\mathbf H_{\nu\nu}, \varrho],
    \label{eq:eom_nu}
\end{align}
where $\varrho$ is the density matrix of neutrinos, and
\begin{equation}
    \mathbf H_{\nu\nu} = \sqrt{2} G_F \int d E'\, d v' (1-v v') (\varrho-\bar\varrho^*).
\end{equation}
Equation~\eqref{eq:eom_nu} also applies to antineutrinos when replacing $\varrho$ by $\bar\varrho$ and $\mathbf H_{\nu\nu}$ by $-\mathbf H_{\nu\nu}^*$.
Assuming the collective unstable mode in the form of $\varrho_{ex}-\bar\varrho_{ex}^* = Q(\Omega,K,E,v)e^{-i(\Omega t-K x)}$,
the off-diagonal part of Eq.~\eqref{eq:eom_nu} becomes
\begin{align}\label{eq:LEQ1}
    & \left[ \Omega - K v - \Phi(v) \right]  Q(E,v) \nonumber\\
    = & -\sqrt{2}G_F G(E,v) \int dE'\, dv' (1- v v') Q(E',v'),
\end{align}
where $\Phi(v)=\sqrt{2}G_F\int dE'\, dv'\, (1-v v') G(E',v')$, and $G(E,v)=\varrho_{ee}-\varrho_{xx}-\bar\varrho_{ee}+\bar\varrho_{xx}$.

At each time $t$ and radius $r$, we solve for the eigenvalues of $\Omega$ and the corresponding eigenvectors for a wide range of $K$'s.
If FFI exists, the unstable mode whose growth rate $\mathrm{Im}(\Omega)$ is maximal among all $K$ modes is recorded.  

Figure~\ref{fig:2d_ImOmega}(a) shows the obtained maximum growth rate $\mathrm{Im}(\Omega)$ as a function of time and radius.
The magnitude of FFI growth rates after the second core bounce can be larger than $10^5~\mathrm{km}^{-1}$, particularly in the vicinity of the PNS surface. 
As the radius increases, the growth rate decreases by around one or two orders of magnitude, but remains much larger than the unstable growth rate during the pre-second-bounce phase due to the shallow crossings originated from the back-scattered neutrinos, indicated by the yellow region with growth rates only about $10^2~\mathrm{km}^{-1}$ at the upper left corner of the same panel.
This analysis quantitatively confirms the existence of FFIs and the associated FFC time scale in phase transition SNe.

\subsection{Estimation on the flavor conversions}
As these FFIs locate near the neutrinospheres, it requires solving the neutrino quantum kinetic equation ($\nu$QKE)~\cite{sigl1993general,vlasenko2014neutrino,volpe2015neutrino,blaschke2016neutrino} or applying effective prescriptions in a hydrodynamic simulation to capture the possibly associated feedback effect~\cite{xiong2024fast} and the actual impact on SN dynamics~\cite{wang2025effect}.
Nonetheless, we estimate the potential effects from FFCs on the change of flavor content of neutrinos by applying the prescription ``power-$1/2$'' in three flavors obtained in our previous studies \cite{xiong2023evaluating} to the local neutrino distributions at all space-time grids where the FFIs are found.

For each energy-integrated distribution $G(v)=\int dE' G(E', v)$, we determine the crossing velocity and evaluate the integrals of $G(v)$ on both sides of the crossing where $G(v)$ is positive and negative.
Assuming the conservation of the local $\nu$ELN, the prescription provides the survival probability for electron flavors as $P_{ee}(v)$, which can be applied to unoscillated number densities $\varrho^0$ and $\bar\varrho^0$ to obtain the post-oscillation angular distributions
\begin{align}
    \varrho_{ee} & = P_{ee} \varrho^0_{ee} + (1-P_{ee}) \frac{\varrho^0_{\mu\mu}+\varrho^0_{\tau\tau}}{2}, \\
    \bar\varrho_{ee} & = P_{ee} \bar\varrho^0_{ee} + (1-P_{ee}) \frac{\bar\varrho^0_{\mu\mu}+\bar\varrho^0_{\tau\tau}}{2},
\end{align} 
in the absence of any difference between muon and tau flavors.

After obtaining the post-oscillation neutrino angular distributions, we integrate it over the energy and angular space to obtain the corresponding number densities $n_{\bar\nu_\alpha}$ for any species $\alpha$.
We show the $\bar\nu_e$ survival ratio defined as $n_{\bar\nu_e}/n_{\bar\nu_e}^0$ and the angle-integrated $\bar\nu_e$ survival probability, $(n_{\bar\nu_e}-n_{\bar\nu_x}^0)/(n_{\bar\nu_e}^0-n_{\bar\nu_x}^0)$, in Fig.~\ref{fig:2d_ImOmega}(b) and (c), respectively.
Here, we choose $\bar\nu_e$ because its luminosity dominates over other neutrino species after the second bounce.
In the first brief phase of FFI around $t_{\rm p2b}=0$, Fig.~\ref{fig:2d_ImOmega}(b) shows that a reduction of more than 30\% in $\bar\nu_e$ number density takes place, with the maximum converted region generally following the moving shock.
For the later phase of FFI at $t_{\rm p2b}\gtrsim 5.5$~ms, significant reduction of up to $\sim 20\%$ in $n_{\bar\nu_e}$ at the PNS surface also occurs.
For both phases, the minimum value of the $\bar\nu_e$ survival probability reaches $\sim 1/3$, indicating that near-complete flavor equipartition is achieved.

The lower values of the survival ratio for $\bar\nu_e$ (or more reduction of $n_{\bar\nu_e}$) in the first phase than in the second is related to the larger asymmetry in $n_{\bar\nu_e}$ and $n_{\bar\nu_x}$ for the former.
The $\bar\nu_e$ number luminosity is nearly twice of the $\bar\nu_x$ one at the peak [$t_{\rm p2b}\sim 3.4$~ms in Fig.~\ref{fig:nu_property}~(a)], while this ratio reduces to $\sim 1.3$ for $t_{\rm p2b}>7$~ms, leading to relatively more conversion from $\bar\nu_x$ to $\bar\nu_e$ that increases the $\bar\nu_e$ survival ratio.
At larger radii, the estimated survival ratios decrease, consistent with higher values of $\bar\nu_e$ survival probabilities shown in Fig.~\ref{fig:2d_ImOmega}(c), due to the angular crossing moving toward $v=-1$ (see Fig.~\ref{fig:ELN}). 
However, we remind the readers that this trend is obtained based on the simulation data without including neutrino oscillations.
If flavor conversions take place at inner radii, the flavor content at outer radii could be affected when the flavor waves propagate outward~\cite{xiong2024fast}.
Therefore, larger amount of flavor conversions outside the PNS surface may still be possible.

For the upper left corner where the instabilities are weak, the corresponding $\bar\nu_e$ survival ratios are typically $\lesssim 0.05\%$, and are not expected to lead to significant consequences to the SN dynamics and observables.

\subsection{Neutrino signals}
The impact of FFC on the neutrino flavor content near the PNS should affect the emerging neutrino energy spectra and the associated signals from QCD phase transition SNe.
As demonstrated in previous studies~\cite{dasgupta2010detecting,pitik2022exploiting,zha2020gravitationalwave,zha2022impact,lin2024detectability}, the strong millisecond neutrino burst in all flavors guarantees its detection in all current and upcoming large-scale neutrino experiments, which offers robust identification of the phase transition nature with a galactic event. 
In this subsection, we examine the potential signature of FFC in such neutrino signals.

We follow the procedure described in \cite{wu2015effects,fischer2018quark} to calculate the expected neutrino event rates from a galactic QCD SN at a distance to the Earth of 10~kpc in Hyper-Kamiokande (HK) \cite{proto2018hyperkamiokande} and in Deep Underground Neutrino Experiment (DUNE) \cite{abi2020deep}, taking the fiducial volume of 200~kton for HK and 40~kton for DUNE.
The event rate for each distinct detection channel is approximately calculated by
\begin{equation}\label{eq:event_rate}
    \frac{dN}{dt}(t) = N_i \sum_\alpha \int_{E_{\rm th}}^\infty F^\oplus_{\nu_\alpha}(E,t) \sigma_{\nu_\alpha}(E) dE,
\end{equation}
where $N_i$ is the number of target, $F^\oplus_{\nu_\alpha}$ is the neutrino flux arriving at the Earth, $E_{\rm th}$ and $\sigma_{\nu_\alpha}$ are the threshold neutrino energy and the cross section for the corresponding interaction taken from \cite{scholberg2012supernova} with the neutrino-nucleus interaction cross section computed in \cite{kolbe2003neutrinonucleus}, respectively.
The event rate for antineutrinos is calculated in a similar way by replacing $\nu$ by $\bar\nu$ in Eq.~\eqref{eq:event_rate}.
We assume 100\% detection efficiency for simplicity and neglect the Earth matter effect.

For different oscillation scenarios that affect $F^\oplus_{\nu_\alpha}$, we examine three distinct cases: the normal ordering (NO), inverted ordering (IO), and complete flavor equipartition.
Given that $n_{\bar\nu_e}/n_{\nu_e}\sim 1$ at where strong FFIs are present, flavor equipartition can be viewed as a good proxy to estimate the impact of FFC on neutrino signals. 
For $\nu_e$ and $\bar\nu_e$, their fluxes ($F^\oplus_{\nu_e}$ and $F^\oplus_{\bar\nu_e}$) in each of these cases are related to the flux without considering any oscillation effect, $F^{\oplus,0}_{\nu_\alpha}$ and $F^{\oplus,0}_{\bar\nu_\alpha}$, by
\begin{align}\label{eq:flux_nue}
    & F^\oplus_{\nu_e} = s_{13}^2 F^{\oplus,0}_{\nu_e}+c_{13}^2 F^{\oplus,0}_{\nu_x} & \mathrm{(NO),} \nonumber\\
    & F^\oplus_{\nu_e} = \frac{1}{3} F^{\oplus,0}_{\nu_e}+\frac{2}{3} F^{\oplus,0}_{\nu_x} & \mathrm{(equi.),} \nonumber\\
    & F^\oplus_{\nu_e} = s_{12}^2 c_{13}^2 F^{\oplus,0}_{\nu_e}+(c_{12}^2 c_{13}^2+ s_{13}^2) F^{\oplus,0}_{\nu_x} & \mathrm{(IO),}
\end{align}
and
\begin{align}\label{eq:flux_nueb}
    & F^\oplus_{\bar\nu_e} = c_{12}^2 c_{13}^2 F^{\oplus,0}_{\bar\nu_e}+(s_{12}^2 c_{13}^2+s_{13}^2) F^{\oplus,0}_{\bar\nu_x} & \mathrm{(NO),} \nonumber\\
    & F^\oplus_{\bar\nu_e} = \frac{1}{3} F^{\oplus,0}_{\bar\nu_e}+\frac{2}{3} F^{\oplus,0}_{\bar\nu_x} & \mathrm{(equi.),} \nonumber\\
    & F^\oplus_{\bar\nu_e} = s_{13}^2 F^{\oplus,0}_{\bar\nu_e}+c_{13}^2 F^{\oplus,0}_{\bar\nu_x} & \mathrm{(IO),}
\end{align}
where $c_{ij}$ and $s_{ij}$ are for $\cos \theta_{ij}$ and $\sin \theta_{ij}$ with the corresponding mixing angle $\theta_{ij}$ values used in \cite{fischer2018quark}.

\begin{figure}[!hbt]
\includegraphics[width=\columnwidth]{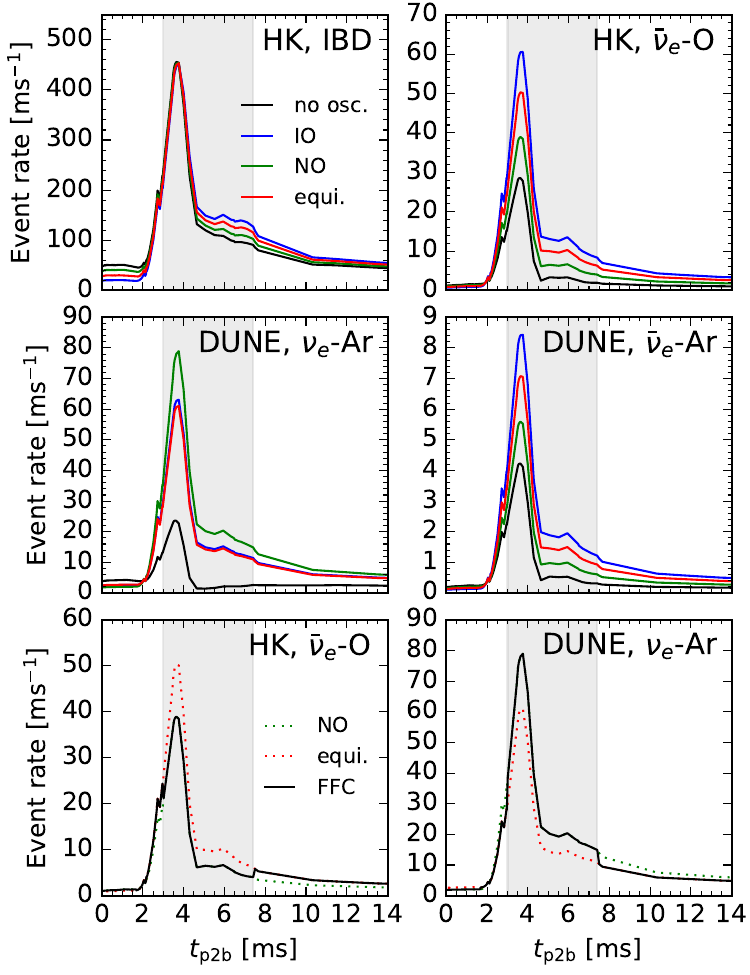}
\caption{\label{fig:eventrate} Event rates as a function of time $t_{\rm p2b}$ in four different channels for the HK (200~kton) and DUNE (40~kton) detectors, respectively, in the top and middle panels.
Comparisons are made among three scenarios for IO, NO, and complete flavor equipartition.
The distance to the galactic QCD SN is 10~kpc.
The bottom panels show the event rates where the complete flavor equipartition is only achieved before 3~ms and after 7.5~ms suggested by the FFC analysis.
A shaded area is marked in all panels where FFC is expected to have no direct impact.}
\end{figure}

\begin{figure}[!hbt]
\includegraphics[width=\columnwidth]{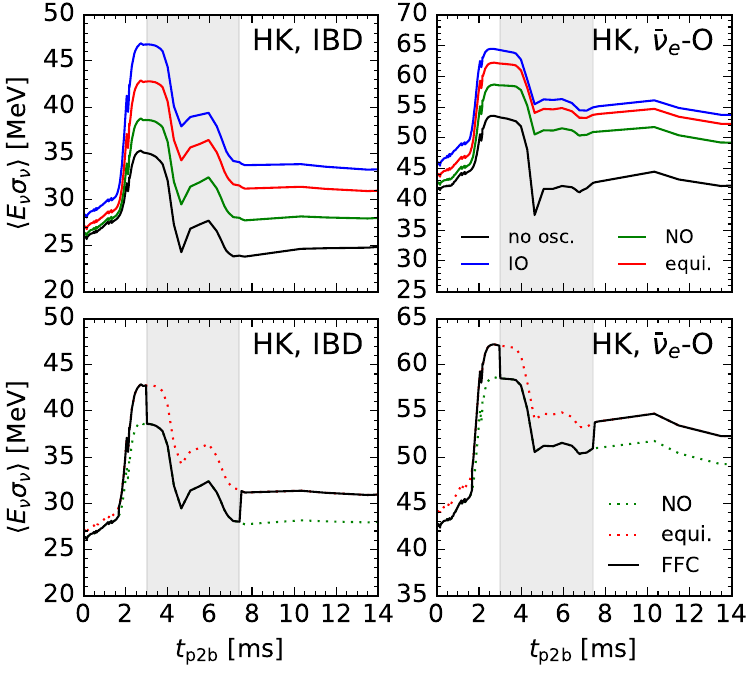}
\caption{\label{fig:eventrate_mean_energy}Mean energy of neutrino events as a function of time $t_{\rm p2b}$ for the HK (200~kton) detectors in IBD and $\bar\nu_e$-$^{16}$O channels, respectively.
The distance to the galactic QCD SN is 10~kpc.
The top panels compare three scenarios for IO, NO, and complete flavor equipartition.
The bottom panels show the event rates where the complete flavor equipartition is only achieved before 3~ms and after 7.5~ms suggested by the FFC analysis.
A shaded area is marked in all panels where FFC is expected to have no direct impact.
}
\end{figure}

The top and middle panels in Fig.~\ref{fig:eventrate} show the expected event rates per millisecond of the inverse $\beta$ decay (IBD), charged-current $\bar\nu_e$-$^{16}$O interaction in HK, as well as the charged-current $\nu_e$ and $\bar\nu_e$ interaction with $^{40}$Ar in DUNE for NO, IO, and flavor equipartition cases.
In addition, we also plot results obtained assuming no flavor conversions at all as references.
Clearly, the event rate from the IBD channel in HK is the largest, giving rise to $\mathcal{O}(10^3)$ events during the 2--3~ms window of the burst.
Interestingly, the event rates of the IBD channel during the peak is nearly independent of the assumed oscillation scenarios, broadly consistent with what was reported in \cite{dasgupta2010detecting,fischer2018quark}. 
Before and after the burst, different oscillation scenarios lead to different IBD event rates.
For all other channels shown in Fig.~\ref{fig:eventrate}, despite their smaller event number than the IBD channel, a clear difference among all scenarios can be observed during and after the burst.

The above results can be understood in terms of $L_\nu^{\rm num}$ and $\langle E_\nu \rangle$ by examining Eq.~\eqref{eq:flux_nueb} since the event rate of a particular channel is roughly proportional to the neutrino flux at Earth multiplied by the square of the difference between their average energy and the threshold energy of the interaction. 
Given that $c_{13}^2$ is close to 1 and $c_{12}^2\sim 2/3$, we expect a smooth transition from $(2F^{\oplus,0}_{\bar\nu_e}+F^{\oplus,0}_{\bar\nu_x})/3$ in NO, to the intermediate scenario with complete flavor equipartition, and finally to $F^\oplus_{\bar\nu_e}\approx F^{\oplus,0}_{\bar\nu_x}$ in IO.
Next, the event rate ratio between NO and IO cases can be approximated by 
\begin{equation}
    \bar R = 
    \left( \frac{1}{3}+\frac{2}{3} \frac{L_{\bar\nu_e}^{\rm num}}{L_{\bar\nu_x}^{\rm num}} \right)
    \left( \frac{1}{3}+\frac{2}{3} \frac{\langle E_{\bar\nu_e} \rangle-E_{\rm th}}{\langle E_{\bar\nu_x} \rangle-E_{\rm th}} \right)^2.
\end{equation}
From Fig.~\ref{fig:nu_property}, we can read $L_{\bar\nu_e}^{\rm num}/L_{\bar\nu_x}^{\rm num}$ is approximately 1.8 at the peak of $t_{\rm p2b}\sim 4$~ms.
If taking $E_{\rm th}\approx 0$ for IBD, we have $(\langle E_{\bar\nu_x} \rangle-E_{\rm th})/(\langle E_{\bar\nu_e} \rangle-E_{\rm th}) \approx 1.33$, which cancels the ratio of number luminosities and leads to $\bar R\approx 1$.
For the channels for $\bar\nu_e$ interactions with O and Ar, $E_{\rm th}\approx 15$~MeV, and hence $(\langle E_{\bar\nu_x} \rangle-E_{\rm th})/(\langle E_{\bar\nu_e} \rangle-E_{\rm th}) \approx 1.9$ as well as $\bar R\approx 0.72$.
This is consistent with the pattern where the event rate at the peak in the NO case is around two third of the IO case.
The event rate with complete flavor equipartition lies in between the NO and IO cases, given that only 1/3 of the unoscillated $F^{\oplus,0}_{\bar\nu_e}$ remains in Eq.~\eqref{eq:flux_nue}.

For the neutrino absorption channel, Eq.~\eqref{eq:flux_nue} implies $F^\oplus_{\nu_e}\approx F^{\oplus,0}_{\nu_x}$ for NO, and the IO scenario behaves similarly to the one with complete flavor equipartition after FFCs.
The event rates for the $\nu_e$-Ar channel in the IO and flavor equipartition cases are roughly three fourth of that in the NO case at the peak.
This can be understood by examining $R$ corresponding to Eq.~\eqref{eq:flux_nue}
\begin{equation}
    R = 
    \left( \frac{2}{3}+\frac{1}{3} \frac{L_{\nu_e}^{\rm num}}{L_{\nu_x}^{\rm num}} \right)
    \left( \frac{2}{3}+\frac{1}{3} \frac{\langle E_{\nu_e} \rangle-E_{\rm th}}{\langle E_{\nu_x} \rangle-E_{\rm th}} \right)^2.
\end{equation}
Given that the luminosity ratio $L_{\nu_e}^{\rm num}/L_{\nu_x}^{\rm num}$ is approximately 0.8, and taking $E_{\rm th}\approx 0$ so that $\langle E_{\nu_x} \rangle/\langle E_{\nu_e} \rangle \approx 1.4$, we have $R\approx 0.76$.

We note that the event rates shown in Fig.~\ref{fig:eventrate} are computed assuming flavor equipartition takes place all the time. 
If we take into account the actual presence of FFIs in different phases as analyzed in previous sections and assume that no other types of flavor instabilities occurs, the corresponding event rates will jump from either the IO or NO curve to the flavor equipartition curve after the peak.
To accurately reflect the FFC analysis, we consider a case transitioning from the equilibration scenario to NO at $t_{\rm p2b}=3$~ms and back to equilibration again at $t_{\rm p2b}=7.5$~ms,  accounting for the traveling time to a radius of 500~km where the luminosity and spectrum are evaluated.
A shaded area is also marked in all panels to indicate the time window where FFC is expected to have no direct impact.
The event rates labeled as ``FFC'' are shown in the bottom panels of Fig.~\ref{fig:eventrate} for two detection channels. 
The first jump of the total event rate at $\sim 3$~ms occurs when the event rates are rising steeply during the burst and appears difficult to be identified. 
On the other hand, an anti-correlated change in event rates for different detection channels can be observed near the second jump.
Such a jump, if detectable in neutrino signals, can also be used to infer the occurrence of FFCs. 
The delay time of this jump from the peak time may also be related to other characteristics of the phase transition such as the progenitor mass as discussed in Sec.~\ref{subsec:progenitor}.

Although the total event number of the IBD channel only weakly depends on the oscillation scenario, difference should exist in the reconstructed neutrino energy spectrum, which could be used as diagnostics for FFCs~\cite{nagakura2023basic,xiong2024fast,abbar2025using}.
For simplicity, we calculate the mean energy of neutrino events
\begin{equation}
    \langle E_\nu \sigma_\nu \rangle = \frac{ \sum_\alpha \int_{E_{\rm th}}^\infty F^\oplus_{\nu_\alpha}(E,t) \sigma_{\nu_\alpha}(E) E dE }{ \sum_\alpha \int_{E_{\rm th}}^\infty F^\oplus_{\nu_\alpha}(E,t) \sigma_{\nu_\alpha}(E) dE },
\end{equation}
as a measure for the change in the neutrino energy spectra.
The left panels in Fig.~\ref{fig:eventrate_mean_energy} show that for the IBD channel, the values of $\langle E_\nu \sigma_\nu \rangle$ in three oscillation scenarios can differ from each other by $\sim 4$~MeV.
Because the mean energy peaks slightly earlier than the luminosity (see Fig.~\ref{fig:nu_property}), a transition from NO to complete flavor equilibration scenario at $t_{\rm p2b}=3$~ms is more pronounced than that in the total event rate before the shaded area of the non-FFC phase.
For the second jump at $7.5$~ms, an increase of the mean energy from $\sim 27$~MeV to 31~MeV appears.
Similar features can also be found for the $\bar\nu_e$-$^{16}$O channel shown in the right panels of Fig.~\ref{fig:eventrate_mean_energy}.

\subsection{Potential implications on nucleosynthesis}
The SNe driven by the hadron-quark phase transition has been shown as a possible $r$-process site, where heavy elements above the second $r$-process peak are being formed in neutron-rich materials of the direct ejecta shortly after the second bounce as well as in the later NDW.
As the yields of heavy elements depend on accurate modeling of the matter composition and hence neutrino flavor content, we explore the impact of FFC discussed above on nucleosynthesis outcome in direct ejecta.

\begin{figure}[!hbt]
\includegraphics[width=\columnwidth]{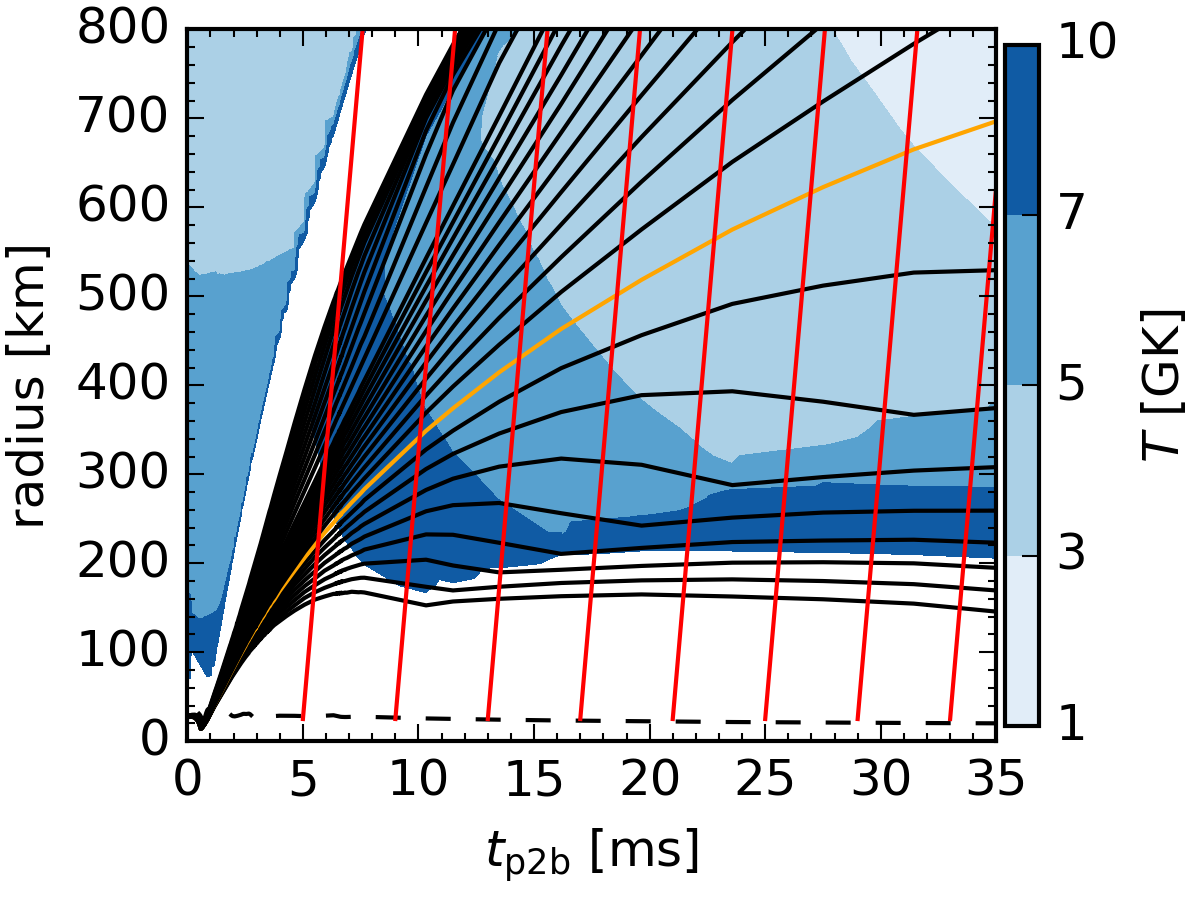}
\caption{\label{fig:2d_potential_effect} Trajectories of Lagrangian mass ejecta (black curves, enclosed mass $1.976 M_\odot\leq M\leq 1.99 M_\odot$) on top of the relevant temperature regions (blue colored domains) with respect to time $t_{\rm p2b}$ and radius $r$. The domain with $T>10$~GK is not colored. Red lines show the trajectories of relativistic neutrinos with the radial velocity $v=1$ emitted from the PNS (black dashed curve). The orange curve shows the selected trajectory for the nucleosynthesis calculation.}
\end{figure}

Since the first phase of FFI mainly follows the position of the shock front, the resulting change of the neutrino content does not affect the direct ejecta behind the shock.
On the other hand, despite the delayed appearance of the second FFI phase of $\sim 5$~ms after the second bounce, neutrinos emitted after that intersect with the trajectories of the direct ejecta, as shown in Fig.~\ref{fig:2d_potential_effect}, and can thus affect the nucleosynthesis condition therein.
For example, neutrinos emitted at $t\sim 5$~ms (first red solid line) intersect the trajectories of direct ejecta (black curves whose radii increase monotonically over time) at $\sim 200$--500~km.
At these radii, neutrino fluxes are still high enough to affect the electron fraction of the ejecta.

We use one trajectory to demonstrate the potential implications for FFCs on the nucleosynthesis consequence.
This trajectory is highlighted by orange color in Fig.~\ref{fig:2d_potential_effect} and has the highest abundance yields beyond second $r$-process peak among all trajectories in direct ejecta.
We introduce two cases to mimic the effect of FFCs and compare them with results obtained without including FFCs. 
The first case assumes, once again, complete flavor equipartition above the neutrinospheres and use the corresponding fluxes and mean energies of all neutrino species to perform the nucleosynthesis calculations.
For the second case, we reduce the fluxes of $\nu_e$ and $\bar\nu_e$ to one third of their original values and keep all other quantities untouched, which corresponds to flavor equipartition if the amount of initial $\nu_x$ is negligible.

For both cases, we use the same nuclear reaction network as in Refs.~\cite{shingles2023selfconsistent,just2023endtoend} to perform the nucleosynthesis calculations.
The network uses experimentally measured nuclear masses, decay rates, and reaction rates, together with theoretical predictions when experimental data are not available (see details in \cite{mendoza2015nuclear}).
We use the set of nuclear reactions with neutron capture and photo-disintegration reactions corrected based on HFB21 mass model \cite{goriely2010further} and the beta decay rates provided by \cite{marketin2016largescale} for neutron-rich nuclei.
The density evolution of the ejecta trajectory is taken directly from the SN simulation before $t_{\rm p2b}$=35~ms. 
After that, we extrapolate the density evolution by $\rho\propto t^{-3}$, featuring homologous and adiabatic expansion.
The temperature evolution of the ejecta takes into account the feedback from nuclear heating and is calculated by the reaction network.
The (anti)neutrino absorptions on nucleons are included because they can be strongly affected by FFCs.

\begin{figure}[!hbt]
\includegraphics[width=\columnwidth]{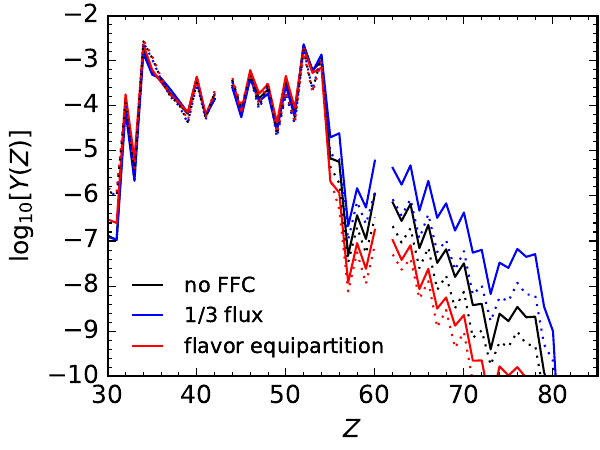}
\caption{\label{fig:nusyn} Comparison of the elemental abundance $Y(Z)$ as a function of atomic number $Z$ from the orange trajectory in Fig.~\ref{fig:2d_potential_effect} (solid curves) and all direct ejecta (dotted curves) without and with including the treatment of FFCs.}
\end{figure}

\begin{figure}[!hbt]
\includegraphics[width=\columnwidth]{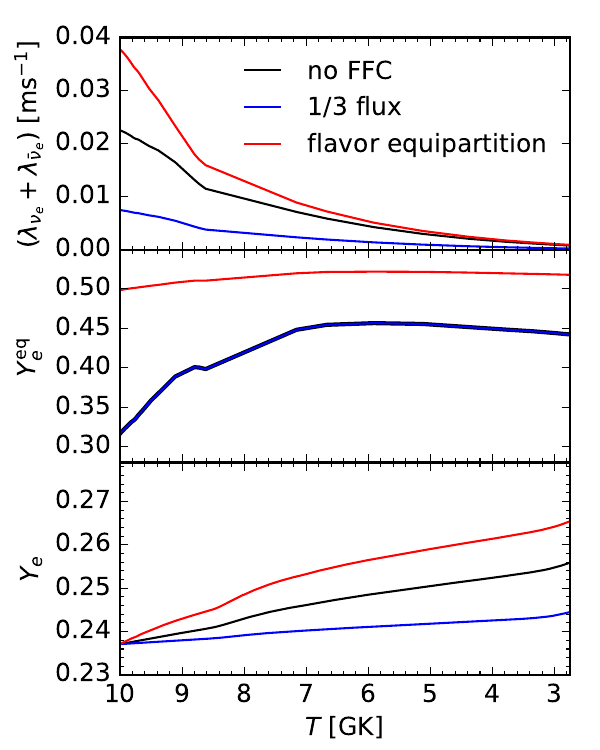}
\caption{\label{fig:nusyn_ye_lambda}Comparison of the total neutrino absorption rate and electron fraction $Y_e$ as a function of temperature $T$ from the orange trajectory in Fig.~\ref{fig:2d_potential_effect} without and with including the treatment of FFCs.}
\end{figure}

The solid lines in Fig.~\ref{fig:nusyn} shows the elemental abundance yields for the example trajectory at 1~Gyr for the cases with and without FFCs.
For the case assuming flavor equipartition due to FFCs, it reduces the amount of heavy element production above the second $r$-process peak by nearly a factor of 10. 
The reasons are twofold.
First, although flavor equipartition results in an overall reduced $\nu_e$ and $\bar\nu_e$ fluxes, it also enhances the $\nu_e$ and $\bar\nu_e$ mean energies.
When the increasing of neutrino mean energy dominates over the flux reduction, the absorption rates becomes larger.
The top panel of Fig.~\ref{fig:nusyn_ye_lambda} compares the sum of $\nu_e$ and $\bar\nu_e$ absorption rates, $\lambda_{\nu_e}+\lambda_{\bar\nu_e}$ among different oscillation scenarios.
The total absorption rate in complete flavor equipartition is about 1.8 times larger than the case without any flavor oscillations at 10~GK.
Although this ratio decreases to $\approx 1.3$ at 8~GK, it remains larger than 1 at $T\gtrsim 5$~GK.
Second, unlike the typical NDW where $Y_e$ at relevant temperatures can be well approximated by the equilibrium value $Y_e^{\rm eq}\equiv (1+\lambda_{\bar\nu_e}/\lambda_{\nu_e})^{-1}$ ignoring the electron captures and neutrino-nucleus interaction, the associated timescale for neutrino absorption is $\sim \mathcal O(100)$~ms, longer than the expansion timescale of the direct ejecta of $\sim \mathcal O(10)$~ms.
Hence, the ejecta $Y_e$ is typically less than $Y_e^{\rm eq}$, and $Y_e^{\rm eq}$ is also slightly less than what is expected in a canonical SN.
For the example without any oscillations, $Y_e^{\rm eq}$ can be as low as 0.3 at 10~GK when the neutrino emission is dominated by $\bar\nu_e$, while it increases to $\approx 0.45$ below 7~GK.
A complete flavor equipartition can enhance the $\nu_e$ absorption rate more strongly and raise the value of $Y_e^{\rm eq}$ to $\approx 0.5$ at 10~GK and $\approx 0.52$ at 7~GK.
Both factors result in a slightly increased $Y_e$ of the ejecta at 3--5~GK by $\approx 0.01$, leading to reduced production of elements above $Z\sim 50$.

For the case where the $\nu_e$ and $\bar\nu_e$ fluxes are reduced to one thirds of their original values, it simply reduces the neutrino absorption rates. 
Because the neutrino mean energy is not changed at all, the value of $Y_e^{\rm eq}$ remains the same.
As a result, the net effect is a reduction of $Y_e$ by $\approx 0.01$ at 3--5~GK, which enhances the abundance yields of heavy elements.

We note here that similar effects happen for other trajectories in the direct ejecta, whose overall nucleosynthesis yields are shown by the dotted curves in Fig.~\ref{fig:nusyn} for all different schemes with or without including FFCs.

\section{Discussion and conclusions}
\label{sec:discussion}
We mainly explored the conditions for the occurrence of FFI in hadron-quark phase-transition SN based on a 25~$M_\odot$ progenitor model and have found their general presence around the PNS surface after the second shock bounce triggered by phase transition. 
By examining the evolution of the characteristic thermodynamic quantities in the period of a few ten milliseconds after the second bounce and the corresponding neutrino emission properties, we delineate their relations with the existence of FFIs.

Due to the protonization of the PNS interior caused by the strong heating of the second bounce shock, the asymmetry factor for neutrino number density $n_{\bar\nu_e}/n_{\nu_e}$, a key indicator of FFIs, switches from smaller than 1 to larger than 1 for most regions above the bounced inner core.
By analyzing the neutrino angular distributions obtained from the SN simulation, we found that ELN angular crossings exist in extended spatial regions around $n_{\bar\nu_e}/n_{\nu_e}\sim 1$ primarily in two distinct phases. 
First, the angular crossings above the PNS appear for a very brief phase of $\sim 1$~ms, which is related to the rapid change of thermodynamical properties during the final millisecond of the collapse and the launch of the shock wave. 
Afterwards, the $\bar\nu_e$-dominated condition due to protonization prohibits ELN crossings for $\sim 5$~ms. 
Crossings reappear in an extended region below and around the PNS surface after that, and robustly persist for several tens of milliseconds.
Stability analysis confirmed that the FFI growth rates in these regions reach $\sim \mathcal{O}(10^5)$~km$^{-1}$, which should lead to vigorous FFCs around the FFI domains.
Moreover, by applying the empirical flavor redistribution formulas to neutrino angular distributions in domains where FFIs are found, we found that flavor equipartition due to FFCs around $n_{\bar\nu_e}/n_{\nu_e}\sim 1$ can be achieved for both phases. 
Qualitatively similar behaviors have also been observed in another phase-transition SN model with a $40~M_\odot$ progenitor mass.

With the identification of the presence of FFIs and the implied flavor equipartition, we have further explored the potential impact of FFCs on neutrino signals and on $r$-process nucleosynthesis in the direct ejecta of phase transition SN. 
For the neutrino signals, we estimated the expected event rates for a galactic phase transition SN at 10~kpc in HK and in DUNE assuming different flavor oscillation scenarios in NO, IO, and flavor equipartition.
Substantial differences generally exist in the channels of IBD, $\bar\nu_e$-O, $\nu_e$-Ar, and $\bar\nu_e$-Ar.
While the total event rate of the IBD channel around the first peak of the millisecond neutrino burst in HK shows less sensitivity to the chosen oscillation scenario, a difference is expected in the reconstructed neutrino energy spectrum.
A detection of a sudden jump in event rates during the post-bounce phase may be used to identify the occurrence of FFCs. 

For the impact on nucleosynthesis, we demonstrated that flavor equipartition results in a slight increase of neutron richness of the direct ejecta, mainly due to increased $\nu_e$ average energy from the conversion of $\nu_x$ to $\nu_e$. 
This effect results in reduced heavy element abundance above the second $r$-process peak by a factor of a few. 
If we assume that FFCs lead to a reduction of $\nu_e$ and $\bar\nu_e$ fluxes to one third of their original values, it results in reduced $Y_e$ and increased heavy element production above the second peak. 

Our work has established the presence of FFI in QCD phase transition SNe and points to a new possibility of investigating neutrino flavor conversions with the associated observables.
Further studies are needed to address effects due to various assumptions taken in this work, such as the break of spherical symmetry, beyond simple post-processing data from a SN model without accounting for feedback of flavor conversions, as well as the dependence on the employed EOS and the variety of progenitor masses.
In addition, the nature of the QCD phase transition in SNe remains inconclusive and could well be of the second-order or crossover type.
For any type other than the first-order, the entire scenario may change qualitatively.
All these steps are essential for obtaining accurate multimessenger signal predictions of phase transition SNe, which may ultimately be used to probe the nature of dense matter at the highest density.

\begin{acknowledgments}
We thank Sajad Abbar for useful conversations.
We also thank the anonymous referee for the swift response and helpful suggestions.
ZX and GMP acknowledge support of the European Research Council (ERC) under the European Union’s Horizon 2020 research and innovation program (ERC Advanced Grant KILONOVA No. 885281), the Deutsche Forschungsgemeinschaft (DFG, German Research Foundation)—Project-ID 279384907—SFB 1245, and MA 4248/3-1.
ZX also acknowledges support of the ERC Starting Grant (NeuTrAE, No. 101165138).
M.R.W. acknowledges support of the National Science and Technology Council, Taiwan under Grant No.~111-2628-M-001-003-MY4, the Academia Sinica under Project No.~AS-IV-114-M04, and Physics Division of the National Center for Theoretical Sciences, Taiwan.
TF and NKL acknowledge the support of the Polish National Science Center (NCN) under grant numbers~2023/49/B/ST9/03941 (TF) and 2023/49/N/ST9/03995 (NKL).
TF also acknowledges support of the Institut Henri Poincaré (UAR 839 CNRS-Sorbonne Université), and LabEx CARMIN (ANR-10-LABX-59-01).
The supernova simulations were performed at the Wroclaw Center for Scientific Computing and Networking (WCSS).
The calculations related to neutrino flavor instability were performed at the Virgo computing cluster of the GSI Helmholtzzentrum {f\"ur} Schwerionenforschung.
We acknowledge the following software: \textsc{matplotlib}~\cite{matplotlib}, \textsc{numpy}~\cite{numpy}, and \textsc{scipy}~\cite{scipy}.

The work is partially funded by the European Union. Views and opinions expressed are however those of the author(s) only and do not necessarily reflect those of the European Union or the European Research Council Executive Agency. Neither the European Union nor the granting authority can be held responsible for them.
\end{acknowledgments}

\vspace{.2in}
The data that support the findings of this article are openly available \cite{dataset_occurrence}.

\bibliographystyle{apsrev4-1}
\bibliography{main.bbl}

\end{document}